\documentclass[notitlepage,amsmath,superscriptaddress,nofootinbib]{revtex4-1}
\usepackage{amssymb,graphicx,braket}

\usepackage{mathpazo}
\usepackage{bm}

\usepackage[usenames,dvipsnames]{color}
\usepackage{soul} 

\newcommand{\D}{\mathrm{d}}
\newcommand{\proj}[1]{\ket{#1}\bra{#1}}
\newcommand{\C}{\mathcal{C}}
\newcommand{\idn}{\hat{\mathbb{I}}}
\DeclareMathOperator{\tr}{Tr}

\usepackage[breaklinks=true]{hyperref}

\hypersetup{
  colorlinks   = true, 
  urlcolor     = blue, 
  linkcolor    = blue, 
  citecolor   =  red 
}

\usepackage{cleveref}
\crefformat{section}{\S#2#1#3}
\crefmultiformat{section}{\S\S#2#1#3}{and~#2#1#3}{, #2#1#3}{, and~#2#1#3}

\begin{document}
\title{Why protective measurement does not establish the reality of
  the quantum state} 
\date{September 29, 2015}

\author{Joshua Combes} 

\affiliation{Institute for Quantum Computing, University of Waterloo,
  Waterloo, Ontario N2L 3G1, Canada} 

\affiliation{Perimeter Institute for Theoretical Physics, 31 Caroline
  St. N, Waterloo, Ontario N2L 2Y5, Canada} 

\author{Christopher Ferrie} 

\affiliation{ Center for Quantum Information and Control, University
  of New Mexico, Albuquerque, New Mexico, 87131-0001, USA}
  
\affiliation{ Centre for Engineered Quantum Systems,
       School of Physics, University of Sydney, Sydney, NSW 2040, Australia}

\author{Matthew S. Leifer} 

\author{Matthew F. Pusey} 

\affiliation{Perimeter Institute for Theoretical Physics, 31 Caroline
  St. N, Waterloo, Ontario N2L 2Y5, Canada}

\begin{abstract}
  ``Protective measurement'' refers to two related schemes for
  finding the expectation value of an observable without disturbing the state of a quantum
  system, given a single copy of the system that is subject to a
  ``protecting'' operation. There have been several claims that these
  schemes support interpreting the quantum state as an objective
  property of a single quantum system. Here we provide three
  counter-arguments, each of which we present in two versions tailored
  to the two different schemes.  Our first argument shows that the
  same resources used in protective measurement can be used to
  reconstruct the quantum state in a different way via process
  tomography.  Our second argument is based on exact analyses of
  special cases of protective measurement, and our final argument is
  to construct explicit ``$\psi$-epistemic'' toy models for protective
  measurement, which strongly suggest that protective measurement
  does not imply the reality of the quantum state. The common theme of
  the three arguments is that almost all of the information comes from
  the ``protection'' operation rather than the quantum state of the
  system, and hence the schemes have no implications for the reality
  of the quantum state.
\end{abstract}
\maketitle

The status of the quantum state is one of the most controversial
issues in the foundations of quantum theory.  Is it \emph{ontic} (a
state of reality) or \emph{epistemic} (a state of knowledge,
information, or belief)?  The likes of de Broglie and Schr{\"o}dinger
initially conceived of the quantum state as a real physical wave,
somewhat akin to a classical field \cite{Bacciagaluppi2009}, whereas
the Copenhagen interpretation views it as a state of knowledge about
the outcomes of future experiments \cite{Faye2008}, more akin to a
classical probability measure than a physical field.  Einstein also
thought that the quantum state represents knowledge \cite{nic}, but,
unlike the Copenhagen school, he thought that this was knowledge about
some deeper underlying reality rather than just the outcomes of
experiments.  In modern parlance, interpretations in which the quantum
state is ontic, in the same sense as a classical field, are dubbed
$\psi$-ontic, and those in which it is epistemic, i.e.\ has the same
status as a classical probability measure, are called
$\psi$-epistemic.  Most current realist interpretations of quantum
theory; such as many-worlds \cite{Everett1957, DeWitt1973,
  Wallace2012}, de Broglie-Bohm theory \cite{Broglie2009, Bohm1952,
  Bohm1952a, Duerr2009}, spontaneous collapse theories
\cite{Ghirardi1986, Bassi2013}, and modal interpretations
\cite{Lombardi2013}; are $\psi$-ontic, but the $\psi$-epistemic view
has recently seen something of a revival in the light of quantum
information theory \cite{Brukner2003, Fuchs2003, toytheory,
  Fuchs2010, Fuchs2010a, Fuchs2013}.  In response to this, the
question of whether Einstein's view; that the quantum state represents
knowledge about a deeper reality; is viable has recently been attacked
rigorously, leading to several theorems, collectively known as
$\psi$-ontology theorems, that aim to show that the quantum state must
be ontic \cite{Pusey2012, Colbeck2012, Colbeck2013a, Aaronson2013,
  Hardy2013, Patra2013a, Mansfield2014, Montina2015} (see
\cite{psireview} for a review).

Protective measurement \cite{protect,meaning} refers to two distinct,
but related, idealized measurement schemes in which a single quantum
system can be probed without changing its state.  The two schemes are
based on the quantum Zeno effect and adiabatic Hamiltonian evolution,
and we call them ``Zeno protected'' and ``Hamiltonian protected''
measurements respectively.  Since protective measurements do not
change the state of the system, they can be used to completely
determine the unknown quantum state of a single quantum system by
performing a tomographically complete sequence of measurements.  Thus,
as concluded in \cite{meaning}, ``this suggests that the wave function
up to a phase may be ontological.''  Such claims have been repeated
\cite{again}, especially in the context of $\psi$-ontology theorems
\cite{shan,lev,hetroh,Gao2015}, where it has been claimed that
protective measurement provides an equally compelling argument for the
reality of the quantum state \cite{shan,Gao2015}.  In this article, we
show that this is not the case.  Protective measurements can easily be
accounted for on the $\psi$-epistemic view.

Ever since protective measurements were first proposed, there has been
much criticism of the claims surrounding them \cite{unruh, rovelli,
  Ghose1995, uffink, dariano, dass, uffink_v_shan, hagar, schcla}.
Much of the early criticism \cite{unruh, rovelli, Ghose1995, uffink,
  dariano} was directed towards the claim that protective measurements
can be used to determine the unknown state of a quantum system.  It is
worth emphasizing that this is not exactly the same issue as whether
or not the quantum state is ontic.  The two issues can only be
identified in operationalist approaches to physical theories, in which
``what is real'' is identified with ``what is measurable''.  However,
it is commonplace in physics to argue for the reality of concepts
indirectly.  For example, the observation of fluctuations in
statistical mechanical systems was regarded as good evidence for the
reality of atoms, long before we developed methods to manipulate and
measure atoms individually.  Thus, in broader realist approaches to
physics, the quantum state may be ontic, and provably so, even if
there is no procedure for measuring it exactly.  This is just what the
recent $\psi$-ontology theorems show, at least under certain
reasonable assumptions.

On the converse side, it might be thought that the ability to measure
something without disturbing the system is at least a sufficient
criterion for its reality.  Along these lines, the
Einstein--Podolsky--Rosen (EPR) criterion for an ``element of
reality'' is \cite{epr}:
\begin{quotation}
  If, without in any way disturbing a system, we can predict with
  certainty (i.e., with probability equal to unity) the value of a
  physical quantity, then there exists an element of physical reality
  corresponding to this physical quantity.
\end{quotation}

However, in the context of protective measurements it is important to
be precise about two further subtleties: \emph{where} the information
about the quantity comes from and \emph{which} degrees of freedom
ought not to be disturbed.

To illustrate the first point, imagine you are handed a quantum system
that is prepared in one of two nonorthogonal states.  A priori, there
is nothing you can do to distinguish them with certainty.  On the
other hand, if, in addition to the system, you are handed a
description of the prepared state written on a piece of paper, then
you can easily distinguish the states without even touching the
system, just by reading what is written on the paper\footnote{A
  similar argument was made by Rovelli \cite{rovelli}.}.  In this
case, there is no doubt that there is an ``element of reality''
corresponding to the state of the system, so EPR are not wrong about
this, but the ``element of reality'' corresponds to the configuration
of pencil marks on the piece of paper, rather than a property of the
system itself.  Nobody would claim that such a procedure has any
implications for whether the quantum state is ontic (i.e.\ whether it
is an intrinsic property of the system itself), since none of the
information about the state actually comes from the system.

In general, if we have access to an additional resource that is
correlated with the quantum state of the system, and the quantum state
can be determined with the help of this resource, then this does not
immediately imply that the quantum state is ontic.  Thus, the question
of whether protective measurement implies the reality of the quantum
state depends on precisely \emph{how much} of the information about
the state of the quantum system comes from the protection operation as
opposed to the system itself.

There is a wide spectrum of possibilities between having no additional
resource and having a complete description of the quantum state, e.g.\
the piece of paper might give some parameters of the state but not all
of them, so we need to determine where the protection operation lies
on this spectrum.  Our first two arguments show that almost all of the
information about the quantity being measured in a protective
measurement comes from the protection operation rather than from the
system itself, which is something that Rovelli \cite{rovelli} and Uffink \cite{uffink} have also argued
for the Hamiltonian case.  In the case of Zeno
protected measurements, it is narrowed down to the expectation values
of the quantity in a set of orthogonal states, and the only
information that comes from this system itself is to pick out one of
these orthogonal states as the one that the system is in.  For
Hamiltonian protected measurements, all of the information comes from
the protection operation and the state of the system is entirely
superfluous.  Thus, we are much closer to the end of the spectrum
where we have a complete description of the state of the system
written on a piece of paper, from which no conclusion that the quantum
state is ontic can be drawn.

Moving on to the second issue, given the role of disturbance in the
EPR criterion, it is no surprise that some of the criticism of
protective measurement has focussed on whether such procedures can
really be implemented without disturbing the system \cite{schcla,
  hagar}.  However, what these critics actually study is the question
of whether protective measurement can be implemented in practice
without modifying the quantum state of the system, but modifying the
quantum state should not be identified with disturbing the system in
general.  Even if protective measurements do not modify the quantum
state, this does not mean that the process does not disturb the
underlying degrees of freedom (otherwise known as the \emph{ontic
  state} of the system), whatever they may be.  For example, in
classical statistical mechanics, the canonical distribution is
invariant under the dynamics of the system, but this does not mean
that the microstate of the system is unchanged.  In a box of gas,
there are a lot of collisions and scattering processes going on, so we
can hardly call the dynamics non-disturbing, even in equilibrium.  It
is only the probability distribution over microstates that is
unchanged, and not the microstate itself.

The point is, if you are investigating the question of whether some
procedure entails the reality of the quantum state, the possibility
that the quantum state is epistemic should be on the table in the
first place, in which case the analogy between quantum states and
probability distributions is appropriate.  If you identify the
question of whether the system is disturbed with the question of
whether its quantum state is modified then this is tantamount to
assuming that the quantum state is identical to the state of reality.
It is difficult to argue against the reality of the quantum state
based on disturbance if you start with a definition of disturbance
that assumes the reality of the quantum state in the first place.

The final argument we present against the idea that protective
measurement implies the reality of the quantum state is a pair of
$\psi$-epistemic toy models that reproduce the salient features of
Zeno and Hamiltonian protected measurements.  In these models the ontic
state is disturbed by the protective measurement even though the
quantum state, which is represented by a probability measure over the
ontic states, is undisturbed in the appropriate limit.  Hence, in
these models, the situation is analogous to the statistical mechanics
example given above, in which the microstate is modified but the
equilibrium distribution is unchanged.  However, in many ways, the
whole issue of disturbance is a bit of a sideshow.  A virtue of the
recent $\psi$-ontology theorems is that they work with a precise
definition of what it means for a theory to be $\psi$-ontic, due to
Harrigan and Spekkens \cite{nic}.  This states that a model is
$\psi$-ontic if the probability measures corresponding to
nonorthogonal quantum states do not overlap, and is $\psi$-epistemic
otherwise (see \cite{psireview} for a detailed discussion of this
definition).  In this sense, we rigorously show that protective
measurements exist in $\psi$-epistemic models.

The remainder of this paper is structured as follows.  In
\S\ref{Review}, we review the basic ideas of protective measurement:
the Zeno case in \S\ref{Review:Zeno} and the Hamiltonian case in
\S\ref{Review:Ham}.  \S\ref{Resource} presents our first argument,
which points out that, given access to the same resources as in
protective measurement, you could determine the quantum state in a
much more straightforward way using quantum process tomography.  In
this procedure, it is clear that most of the information is coming
from the protection operation, so this sharpens the intuition that the
same may be happening in protective measurement.  However, to prove
this rigorously, we need to analyse protective measurement itself,
rather than an alternative procedure, and find a way of determining
which information comes from the protection operation and which from
the system itself.  This we do in our second argument in
\S\ref{Exact}.  In \S\ref{Exact:Zeno}, we analyse a Zeno protected
measurement of a two-outcome observable and derive the Positive
Operator Valued Measure (POVM) corresponding to the whole procedure.
The POVM elements contain all the information in the measurement
procedure that is independent of the state of the system, and hence
they only depend on the protection operation.  We find that the POVM
elements already contain the expectation values of the measured
quantity for an orthogonal set of states, and the only role of the
state of the system is to determine which of these orthogonal states
the system is in.  In \S\ref{Exact:Ham}, we analyse Hamiltonian
protected measurements of quadrature observables made on Gaussian
states.  These we can solve exactly in the Heisenberg picture, which
again cleanly separates the dependence on the system from the
dependence on the protection operation, as the Heisenberg operators
are independent of the initial state of the system.  Here, we find
that all of the information about the quadrature expectation value is
already contained in the Heisenberg operators and is completely
independent of the state of the system.  We also find that, for this
class of states and measurements, procedures equivalent to protective
measurement can be completed in finite time with finite interaction
strength, which shows that criticisms of protective measurement based
on practical considerations are misdirected.  In \S\ref{Toy}, we give
our final argument, which is to construct explicit $\psi$-epistemic
models of protective measurement.  Our Zeno model, discussed in
\S\ref{Toy:Zeno} does not reproduce the quantum predictions exactly,
but it does reproduce the salient features that have been thought to
imply the reality of the quantum state.  For the Hamiltonian case, in
\S\ref{Toy:Ham}, we exploit the existing $\psi$-epistemic model for
Gaussian quantum theory to exactly reproduce the quadrature
measurements considered in \S\ref{Exact:Ham}.  \S\ref{Conc} concludes
and discusses open questions.

Note that throughout we adopt natural units wherein $\hbar = 1$.

\section{Review of protective measurements}

\label{Review}

Here we review the basic ideas of protective measurement.  The paper
\cite{meaning} introduced two methods for protecting the state of a
quantum system during the course of a measurement, Zeno protected
measurement and Hamiltonian protected measurement.  Although the two
schemes led the authors to the same conclusions regarding the meaning
of the wave function, and the overall forms of our counterarguments apply to both, our detailed analyses of the two schemes are quite
different, so it is worth reviewing both of them here.

\subsection{Zeno protected measurements}

\label{Review:Zeno}

To describe Zeno protected measurements, it is helpful to introduce
the usual colourful characters Alice and Bob.  Alice is the person who
is trying to determine the quantum state of the system.  Bob is the
person who initially prepares the system, and it is also his job to
act as godfather to the state of the system, protecting it from
changing throughout the course of Alice's measurements\footnote{We
  dare not ask how much money Alice has to pay Bob for this protection
  racket.}.

\begin{figure}[htb]
  \includegraphics[width=0.8\columnwidth]{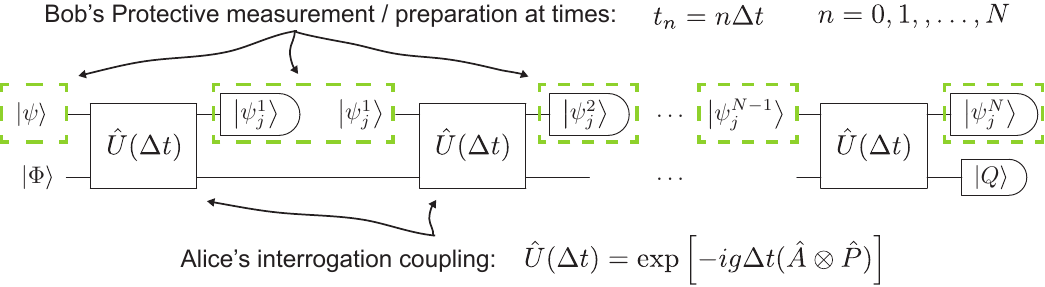}
  \caption{\label{fig1}A quantum circuit schematic of protective
    measurement. The top wire represents the protected system that
    Bob prepares and protects.  Bob's initial preparation $\Ket{\psi}$
    is an element of the basis $\{\ket{\psi_j}\}$.  Note that the
    protecting measurement at times $t_n$ will project the protected
    system onto the state $\ket{\psi_j^n}$. Bob's protection is
    successful if $\ket{\psi_j^n}=\Ket{\psi}$ for all $n$. Alice
    prepares the state $\Ket{\Phi}$ then continuously couples it to
    the protected system, but her coupling is puctuated by Bob's
    interventions. As such her interactions become discretized.
    Finally she measures in the basis $\{\Ket{Q}\}$}
\end{figure}

The intuition behind Zeno protected measurements is as follows: if Bob
makes a \emph{projective} (orthogonal basis) measurement on the system
faster than any other process affecting the system, the state
will---to first order---remain fixed.  This is the quantum Zeno effect
\cite{zeno}.  Whilst this is happening, Alice can measure a complete
set of observables and determine the state of the system.  Moreover,
Alice need not know \emph{which} basis measurement Bob is making, only
that such a measurement \emph{is} being made (see Fig.~\ref{fig1} for
a schematic of this procedure).  In other words, Alice can completely
determine the unknown quantum state of a single \emph{protected}
system.

In a bit more detail, the procedure works as follows.  Bob initially
prepares a quantum system in the state $\Ket{\psi}$, which is known to
him but unknown to Alice, and hands it to Alice at time $t = 0$.  For
simplicity, we assume that the system has zero internal Hamiltonian.
Alice measures observables via the usual von Neumann measurement
coupling scheme.  At time $t = 0$, she prepares a pointer system in a
well-localized state.  For example, the pointer could be a particle in
one-dimension with canonical operators $(\hat Q, \hat P)$ prepared in
a Gaussian state with small uncertainty in $Q$.  Usually, the particle
is assumed to have a very large mass, so that we can ignore the
kinetic term in its Hamiltonian.  If Alice wants to measure the
observable $\hat A$, she couples $\hat A$ to the momentum of the
pointer via the Hamiltonian $\hat H_I = g \hat A \otimes \hat P$ from
time $t = 0$ to $t=1/g$, where $g$ is a coupling constant.  If the
system were not being protected by Bob, then this would be a way of
implementing a conventional von Neumann measurement of $\hat A$, i.e.\
if Alice were to measure the position of the pointer at time $t=1/g$
then she would be very likely to find it close to one of the
eigenvalues of $\hat A$ with probabilities given approximately by the
Born rule, and this becomes exact as the initial uncertainty in $Q$ is
decreased to zero. In Fig.~\ref{fig1} Alice's pointer system is the
lower wire of the quantum circuit.

However, in actual fact, during the course of Alice's measurement, Bob
is doing his best to prevent the state of the system from changing.
At times $t_n = n \Delta t$, where $\Delta t = 1/gN$ and
$n = 0,1,2,\ldots,N$, he sneaks into Alice's lab and instantaneously
measures the system in a basis $\{\Ket{\psi_j}\}$ that includes
$\Ket{\psi}$ as one of the basis elements.  If any pair of his
measurement outcomes differ, then his protection has failed and the
whole procedure is aborted.  However, in the limit
$N \rightarrow \infty$, the probability of this happening tends to
zero (we give the details of this calculation later).  In this limit,
it can also be shown that the wavefunction of the pointer simply
shifts by $\langle \hat A \rangle = \Bra{\psi} \hat A \Ket{\psi}$, so
Alice can read off the expectation value of $\hat A$ by measuring the
pointer position.

If this whole procedure is repeated for a tomographically complete set
of observables, then Alice can determine the initial state of the
system exactly.

\subsection{Hamiltonian protected measurements}

\label{Review:Ham}

In Hamiltonian protected measurements, Alice's measurement procedure
is the same as in the Zeno case.  The only thing that is different is
the way that Bob protects the system.  We assume that Bob has the
ability to set the Hamiltonian of the quantum system, e.g.\ by tuning
an external magnetic field if it is a spin system.  He initially
prepares the system in the state $\Ket{\psi}$ and sets the Hamiltonian
$H_S$ such that $\Ket{\psi}$ is its nondegenerate ground state with
finite gap $\Delta E$ to the first excited state\footnote{Any
  nondegenerate eigenstate with finite gaps to the neighbouring states
  would work just as well, but we use the ground state here for
  simplicity.}.

In the limit $g \rightarrow 0$, we can treat $\hat H_I$ as a small
perturbation.  By the adiabatic theorem, the system and measurement
pointer will remain in the ground state of
$\hat H = \hat H_S + \hat H_I$ throughout the whole procedure.  Since
$\hat H_I = 0$ for $t < 0$ and $t > 1/g$, the system will be in the
ground state of $\hat H_S$ after the measurement is
completed\footnote{One might be concerned that the discontinuous
  change from $\hat H = \hat H_S$ to
  $\hat H = \hat H_S + g \hat A \otimes \hat P$ at $t=0$ and back
  again at $t = 1/g$ violates the assumptions of the adiabatic
  theorem. However, we can instead use the measurement interaction
  $\hat H_I = g(t) \hat A \otimes \hat P$ where $g(t)$ is a smoothly
  varying function with $\int_{t=0}^{t=T} g(t) \, \mathrm{d} t = 1$
  and where $g(t) = 0$ for $t<0$ and $t > T$.}.  As in the Zeno case,
it can again be shown that, in the limit, the pointer shifts by
$\langle \hat A \rangle = \Bra{\psi} \hat A \Ket{\psi}$, so the
expectation value can be read off from the pointer, and the whole
procedure can be repeated for a tomographically complete set of
observables in order to determine $\Ket{\psi}$.

\section{Resource counting argument}

\label{Resource}

Our first argument is to examine the resources available to Alice in
her quest to determine the state of her quantum system.  Firstly, she
has a single copy of the system in an unknown state $\Ket{\psi}$.  If
this were all she had then, of course, she would be unable to
determine the state.  However, there is also Bob's protection
operation, which is either a measurement in an orthonormal basis in
the Zeno scheme, or the ability to evolve the system according to the
unitary $\hat U(t) = e^{-i\hat H_St}$ for an arbitrary time $t$ in the
Hamiltonian scheme.  Both of these operations are correlated to the
unknown state of the system and are used an arbitrarily large number
of times in the protective measurement procedure, i.e.\ in the Zeno
case, the projective measurement is made an arbitrarily large number
of times and in the Hamiltonian case $t$ is arbitrarily large.  Given
this, it is natural to suspect that most of the information about the
unknown state $\Ket{\psi}$ is coming from the protection operation
rather than from the system itself.

To sharpen this intuition, we argue that if Alice is given the exact
same resources that she has available in the protective measurement
schemes then she can determine the unknown state of the system in a
much more straightforward way.  Suppose that instead of Bob applying
the protection operation in the way prescribed by protective
measurement, Alice has black-box access to it.

In the Zeno case, this means Alice has access to a black box that
performs a measurement in a basis $\{\Ket{\psi_j}\}$, where
$\Ket{\psi}$ is one of the basis elements, but Alice does not get to
see the outcome of the measurement.  Thus, from her point of view, it
is a non-selective measurement that implements the quantum channel
\begin{equation} 
  \C(\hat \rho) = \sum_j \proj{\psi_j} \hat \rho \proj{\psi_j}. \label{cdef}
\end{equation}
Alice can use this black box as many times as she likes and pass
whatever systems she likes through it.

Given these resources, Alice can determine $\Ket{\psi}$ in a quite
straightforward way.  First, she puts the system that Bob prepared in
state $\Ket{\psi}$ to one side for later use.  Then, using different
systems that she prepares in known states, she performs process
tomography \cite{processtomo} on her black-box to determine $\C$.
Specifically, if she prepares systems in states $\Ket{\phi_j}$, passes
them through the black box, and performs measurements in the bases
$\{\Ket{\xi^{(k)}_m}\}_m$, then so long as the projectors
$\proj{\phi_j}$ span the vector space of linear operators and the
projectors $\proj{\xi^{(k)}_m}$ also span this space, then the
probabilities
\begin{equation}
  p(m|j,k) = \Bra{\xi^{(k)}_m} \C(\proj{\phi_j}) \Ket{\xi^{(k)}_m},
\end{equation}
determine $\C$ uniquely.  Thus, by
repeating this process a large number of times, she can estimate the
probabilities $p(m|j,k)$, and hence $\C$, to arbitrary accuracy.  Due
to finite sample errors, this will never be exact, but, just as in
Zeno protected measurement scheme, $\C$ can be used an arbitrarily
large number of times, so both procedures only work exactly in the
limit of an infinite number of uses of $\C$.

Having determined $\C$, Alice can calculate the basis
$\{\Ket{\psi_j}\}$ \footnote{This is not completely straightforward as
  Alice only knows $\C$ as a linear map and not the specific
  decomposition in terms of the projectors $\proj{\psi_j}$ given in
  \cref{cdef}.  However, the fixed point set of $\C$ is the set of
  operators that are diagonal in the $\{\Ket{\psi_j}\}$ basis, and
  there are several methods for determining the fixed point set of a
  completely-positive trace-preserving map, e.g. \cite{kribs}.}.  Knowing
this, she now only needs to know which of the basis states the system
was prepared in.  However, since she put the system that Bob gave her
to one side at the beginning, she can now simply measure it in the
basis $\{\Ket{\psi_j}\}$.  She will get the outcome corresponding to
$\Ket{\psi}$ with certainty, and thus will have determined the state
of the system exactly.

Were Alice to use this procedure in the original scenario, it would
look very different to Bob because he will see many different outcomes
of his measurement instead of almost always seeing $\ket{\psi}$. For
some purposes this difference may be crucial. For example if Bob is a
bank unwittingly offering the protection as part of a quantum money
scheme \cite{money}, then while protective measurement can facilitate
successful counterfeiting \cite{forgery}, our alternative would
quickly land Alice in jail. But for understanding what information the
protection process provides, the calculations in the following
sections lead us to believe this difference is not significant.

For the Hamiltonian case, the situation is similar. Alice's black box
now implements the unitary evolution $\hat U(t) = e^{-i \hat H_St}$
and has a setting that allows Alice to vary the time duration $t$.
For a fixed value of $t$, Alice can use process tomography
\cite{processtomo, unitarytomo} to determine the unitary operator
$\hat U(t)$.  This allows her to determine the eigenvalues
$e^{-iE_j t}$ and associated eigenspaces of $\hat U(t)$, where the
$E_j$'s are the eigenvalues of $\hat H_S$.  This does not allow her to
determine $\hat H_S$ uniquely because $e^{-iE_j t}$ is a periodic
function of $E_j$ and there may be degeneracies in the spectrum of
$\hat U(t)$ that are not present in $\hat H_S$, e.g.\ if
$t = 2n\pi/(E_j - E_k)$ for some integer $n$ then
$e^{-iE_j t} = e^{-iE_kt}$ so there is a degeneracy in $\hat U(t)$
even if $E_j \neq E_k$.  Nonetheless, by running this procedure for
different values of $t$, Alice can eventually determine $\hat H_S$.
There are a variety of methods for doing this, e.g. \cite{qec}, and we
omit the details here.

Knowing $\hat H_S$, Alice can determine its ground state and then she
knows that this is $\Ket{\psi}$ without even touching the
system\footnote{If an arbitrary nondegenerate eigenstate is used
  instead of the ground state, Alice must in addition measure
  $\hat H_S$ on the system to determine $\Ket{\psi}$ with certainty.}.

We are not claiming that either of these procedures are models of what
is actually going on in protective measurement. They are just
designed to show that it is unsurprising that Alice can determine the
unknown state given the resources she has available.  In our
procedures, it is clear that the majority of the information about the
unknown state is coming from the protection operation rather than the
system itself.  In the Zeno case, without even touching the system,
the state is narrowed down to one of an orthogonal set, which can then
be determined by a conventional projective measurement.  Only this
information is coming from the system itself.  The Hamiltonian case is
even simpler as the state is determined without ever touching the
system at all.  The fact that this can be done with the same resources
as protective measurement lends credence to the idea that the same
thing may be going on in protective measurements themselves.  It is
not a rigorous argument that this must be so, nor is it a proof that
protective measurement does not imply the reality of the quantum
state.  For that, we shall have to analyse the details of the
protective measurement schemes themselves, which we do in the
following sections.  Nevertheless, we think that this argument casts
doubt on the naive inference from ``the quantum state can be measured
by protective measurements'' to ``the quantum state must be real''.

\section{Exact analyses}

\label{Exact}

The previous argument made it plausible that most of the information
about the unknown state in a protective measurement is coming from the
protection operation rather than from the system itself.  To
demonstrate that this is indeed the case, we need to analyse
protective measurement itself rather than alternative procedures that
use the same resources.  

For the Zeno scheme, we do this for the case of a protective
measurement of a two-outcome observable $\hat{A}$ with eigenvalues
$\pm 1$.  Note that this is sufficient for determining the
unknown state of any finite-dimensional system. For example, on a qubit the Pauli operators $\hat{X}$, $\hat{Y}$ and
$\hat{Z}$, along with the identity operator, span the vector space of
linear operators on the system, and are hence tomographically
complete.

Any measurement procedure acting on a quantum system can be
represented as a Positive Operator Valued Measure (POVM).  In
particular, as pointed out by D'Ariano and Yuen \cite{dariano}, this
still applies to procedures such as protective measurement that
consist of an arbitrarily long sequence of steps.  In the case of
protective measurement, the POVM depends on the unknown state of the
system because the protection operation is correlated with it.
However, by writing things in this way we can cleanly separate the
information that comes from the protection operation from the
information that comes from the quantum system.  Parameters that
appear in the POVM come from the protection operation because they are
independent of what the state of the system is, i.e.\ if Bob were to
give Alice a system prepared in the state $\Ket{\phi}$ but protect it
with a measurement designed to preserve $\Ket{\psi}$ then the POVM
would be the same as if Bob gave Alice the correct state $\Ket{\psi}$.
We find that, as indicated by the previous argument, almost all of the
information about $\Ket{\psi}$ is already present in the POVM.  It
contains complete information about the expectation values
$\Bra{\psi_j}\hat A \Ket{\psi_j}$ for the basis states $\Ket{\psi_j}$
of the protection operation, and the only role of the system is to
select which of these applies.

The Hamiltonian case is more tricky to analyse exactly due to its
reliance on the adiabatic theorem.  However, we can perform an exact
analysis for the protective measurement of quadrature observables
where the protected state is one of a set of overlapping continuous
variable Gaussian states.  Again, this is sufficient to determine the
unknown state, as the states we consider can be distinguished by their quadrature expectation values.

We describe the dynamics in the Heisenberg picture, which provides
another way to separate the dependence on the protection operation
from the dependence on the unknown state, and, in this case, we find
that the information learned by Alice comes entirely from the
protection operation.  We also find that there are procedures that
yield equivalent information to protective measurement that can be
performed with perfect accuracy in finite time, which shows that
criticisms of Hamiltonian protected measurements based on practical
considerations are misdirected.

\subsection{Zeno protected measurements}

\label{Exact:Zeno}
 
In this section, we analyse the effect of a Zeno protected measurement
of a two outcome observable $\hat{A}$ with eigenvalues $\pm 1$
that is designed to preserve the state $\Ket{\psi}$.  We analyse
the action of this procedure when it is applied to an arbitrary
initial state of the system $\hat \rho$, i.e.\ not necessarily
$\Ket{\psi}$.

Alice performs her measurement of $\hat{A}$ by coupling the system to
a one dimensional pointer with canonical operators $(\hat Q, \hat P)$
via the Hamiltonian
\begin{align}
\hat H_I = g \hat A \otimes  \hat P
\end{align}
where $g$ is a coupling parameter.  This Hamiltonian acts from $t_0 = 0$
to $t_N = N \Delta t$ where $\Delta t =1/g N $. 

Bob's protection is via $N+1$ projective measurements of the system in
the $\{\Ket{\psi_j}\}$ basis, where $\Ket{\psi}$ is one of the basis
elements, evenly spaced at times $t_n=n \Delta t$ for $n = 0,1,\ldots,N$. If
Bob gets different outcomes for any of his measurements then he aborts
the procedure.

Let $\Pi_{\pm}$ be the projectors onto the $\pm 1$ eigenspaces of $\hat A$, so that $\hat{A} = \Pi_+ - \Pi_-$.

Between each protection operation the evolution is given by
\begin{equation}
\hat U(\Delta t)=   \exp[-(i/N) \hat H_I] = \exp[-(i/N) \hat A\otimes  \hat P  ] = \exp[-(i/N) (\hat\Pi_+\otimes  \hat P - \hat\Pi_-\otimes  \hat P) ] .
\end{equation}
Because $[\hat\Pi_+\otimes  \hat P , \hat\Pi_-\otimes  \hat P]=0$ we can simplify this to
\begin{align}
 \hat U(\Delta t)&=  \exp [-(i/N) \hat\Pi_+\otimes \hat P]\exp[ +(i/N)  \hat\Pi_-\otimes \hat P ] 
= \hat\Pi_+\otimes \exp[-(i/N) \hat P]+ \hat\Pi_-\otimes \exp[+(i/N)\hat P] 
\end{align}

Suppose the pointer is initially prepared in the Gaussian state
$\ket{\Phi} = \int_{-\infty}^{\infty} \Phi(Q)\ket{Q}dQ$, where
\begin{equation}
  \Phi(Q) = \frac{1}{(\pi\sigma^2)^{1/4}} \exp\left(
    -\frac{Q^2}{2\sigma^2} \right),
\end{equation}
and is measured in the $\{\ket{Q}\}$ basis at the end of the
procedure. 

The effect of this procedure can be viewed as a generalized
measurement on the system that depends on $\Ket{\psi}$ via the
protection operation.  Specifically, if the outcome of the final
measurement is $Q$ then the state of the system is updated via
\begin{equation}
  \hat{\rho} \rightarrow \frac{\hat M_Q \hat{\rho} \hat M_Q^{\dagger}}{\tr (\hat E_Q \hat{\rho})},
\end{equation}
where $\hat M_Q$ is a Kraus operator, which we shall
specify shortly, and $\{\hat E_Q\}$ is a POVM given by
$\hat E_Q = \hat M_Q^{\dagger} \hat M_Q$.  The probability of getting
the $Q$ outcome is given by $\tr (\hat E_Q \hat \rho)$.  Since
$\hat E_Q$ is independent of the initial state of the system, any
information it contains about $\Ket{\psi}$ comes from the protection
operation rather than the system.  The system itself contributes
information only to the extent that $\tr (\hat E_Q \hat \rho)$ depends
on $\hat \rho$.

\begin{figure}
  \includegraphics[scale=0.75]{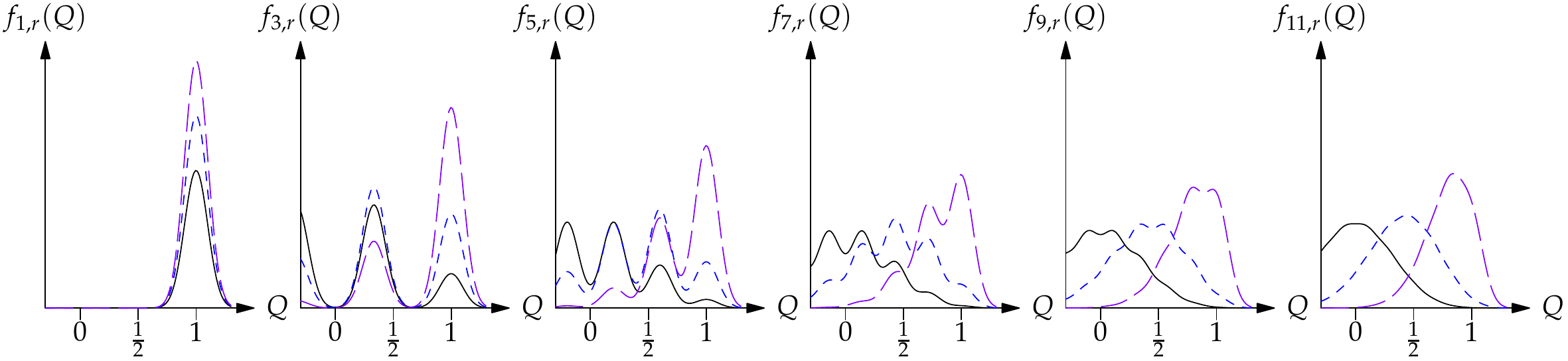}
  \caption{$f_{N,r}$ for $N = 1, 3, 5, 7, 9, 11$, $r = 0.5, 0.7, 0.9$ (black, blue short dashes and purple long dashes respectively), and $\sigma = 0.1$. For these parameters the Gaussian approximation of \cref{fapprox} can be visualised at around $N\ge11$.}
  \label{fnp}
\end{figure}

The generalized measurement operators are given by
\begin{align}
M_Q 
&= \sum_j [\bra{\psi_j} \otimes \bra{Q}]\hat U(\Delta t)
  [\proj{\psi_j} \otimes \idn] \hat U(\Delta t) [\proj{\psi_j} \otimes
  \idn] \dotsm \hat U(\Delta t) [\ket{\psi_j}\otimes\ket{\Phi}] \\
&= \sum_j \proj{\psi_j}\bra{Q} \big \{ \braket{\psi_j|\hat
  \Pi_+|\psi_j} \exp[(-i/N) \hat P]+ \braket{\psi_j|\hat
  \Pi_-|\psi_j}\exp[+(i/N) \hat P] \big \}^N \ket{\Phi} 
\end{align}
note that $\braket{\psi_j|\hat \Pi_\pm|\psi_j}$ is a real number and the operator exponentials commute. Lets introduce some simplified notation $r_j = \braket{\psi_j|\hat \Pi_+|\psi_j}$ and $1-r_j = \braket{\psi_j|\hat \Pi_-|\psi_j}$.
Now we have
\begin{align}
M_Q 
&= \sum_j \proj{\psi_j} \left  \{  \sum_{n=0}^N \binom{N}{n} \bra{Q} r_j^n (1-r_j)^{N-n} e^{-in\hat P/N}e^{+i(N-n)\hat P/N}\ket{\Phi} \right \}\\
&= \sum_j \proj{\psi_j} \left  \{  \sum_{n=0}^N \binom{N}{n} \bra{Q} r_j^n (1-r_j)^{N-n} e^{-i (2n-N)\hat P/N}\ket{\Phi} \right \}\\
&= \sum_j f_{N,r_j}(Q)\proj{\psi_j} 
\end{align}
where we define $f_{N,r}(Q)$, plotted in Fig.~\ref{fnp}, as
\begin{align}
  f_{N,r}(Q) 
  &= \sum_{n=0}^N \binom{N}{n} r^n (1-r)^{N-n}  \bra{Q} e^{-i (2n-N) \hat P/N}\ket{\Phi}\\
  &= \sum_{n=0}^N \binom{N}{n} r^n (1-r)^{N-n}  \Phi[Q-(2n-N)/N]
\end{align}

The corresponding POVM element is
\begin{equation}
\hat  E_Q =\hat M_Q^\dagger \hat M_Q = \sum_j f_{N,r_j}^2(Q)  \proj{\psi_j} .
\end{equation}
It is already evident that the POVM contains information about
$\Ket{\psi}$.  All the POVM elements are diagonal in the
$\{\Ket{\psi_j}\}$ basis and the functions $f_{N,r_j}$ depend on the
operator $\hat{A}$ via $r_j= \braket{\psi_j|\hat\Pi_+|\psi_j}$.  If
the system happens to be prepared in the state $\Ket{\psi}$ then this
simply serves to pick out the $f_{N,\braket{\psi|\hat\Pi|\psi}}^2$
term.  It must therefore be the case that all of the information about
the expectation value $\braket{\psi |\hat A | \psi}$ that is
observed in the protective measurement is already present in the POVM
via the $f_{N,\braket{\psi|\hat\Pi_+|\psi}}^2$ term, which comes from
the protection operation, and the initial state just serves to pick
out $\Ket{\psi}$ from amongst the orthogonal possibilities
$\{\Ket{\psi_j}\}$.

To see that this is the case, we examine the large $N$ limit.  For
large $N$ we can approximate the binomial distribution by the normal
distribution:
\begin{equation}
  \binom{N}{n}r^n(1-r)^{N-n} \approx \frac{1}{\sqrt{2\pi
      Nr(1-r)}}\exp\left(-\frac{(n-Nr)^2}{2Nr(1-r)}\right). 
\end{equation}

This gives
\begin{align}
  f_{N,r}(Q)& \approx \frac{1}{\sqrt{2\pi Nr(1-r)}} \int_{-\infty}^{\infty} dn \Phi[Q-(2n-N)/N] \exp\left(-\frac{(n-Nr)^2}{2Nr(1-r)}\right) \\
  & \approx\frac{\sqrt{\sigma}}{\pi^{1/4}\gamma_{N,r}} \exp\left(-\frac{[Q-(2r -1)]^2}{2\gamma_{N,r}^2}\right)\label{fapprox}
\end{align}
(on the second line we performed the integral) where
\begin{equation}
  \gamma_{n,p} = \sqrt{\frac{4r(1-r)}{N} + \sigma^2}.
\end{equation}
For very large $N$, $\gamma_{N,r} \approx \sigma$ and then $f_{N,r}(Q) \approx   \Phi[Q-(2r -1)]$, giving 
\begin{align}
\hat E_Q \approx \sum_j \proj{\psi_j}   \Phi^2[Q-(2r_j -1)].
\end{align} 
So in this limit, the $\proj{\psi_j}$ term of $E_Q$ is a Gaussian
centered about the expectation value
$\braket{\psi_j|\hat{A}|\psi_j}=(2r_j -1)$.\footnote{If we also take the limit $\sigma \to 0$, so that $\Phi$ is a Dirac delta, then the whole procedure amounts to a projective measurement of the observable $\sum_j (2r_j - 1)\ket{\psi_j}\bra{\psi_j}$ analogous to equation 14 of \cite{uffink}.} This can be seen by noting
that $r_j = \braket{\psi_j|\hat \Pi_+|\psi_j}$ and
$1-r_j = \braket{\psi_j|\hat \Pi_-|\psi_j}$ can be related to the expected
value of the observable $\hat{A}$ via
$\braket{\hat{A}} = \braket{\hat \Pi_+}-\braket{\hat \Pi_-}=r_j -(1-r_j) = 2r_j -1$. From
this, we see that the pointer shifts for each of the basis states
$\Ket{\psi_j}$ are already completely determined by the POVM, and the
system simply serves to determine which of these shifts is applied.

Finally, for completeness, the case where Bob aborts the procedure
corresponds to the POVM element
\begin{equation}
  \hat E_\text{abort} = \hat I - \int_{-\infty}^\infty \hat E_Q dQ.
\end{equation}
Using the approximation \cref{fapprox} we have
\begin{equation}
  \hat E_\text{abort} \approx \sum_j \proj{\psi_j} \left(1 - \frac{\sigma}{\sqrt{\pi}\gamma_{N,r}^2}\int_{-\infty}^{\infty}\exp\left(-\frac{[Q-(2r -1)]^2}{\gamma_{N,r}^2}\right)dQ\right) = \sum_j \proj{\psi_j} \left(1 - \frac{\sigma}{\gamma_{N,r_j}}\right),
\end{equation}
which tends to zero as $N$ tends to infinity. 

\subsection{Hamiltonian protected measurements}

\label{Exact:Ham}

In the Hamiltonian case, we can again separate the dependence on the
protection operation from the dependence on the initial state of the
system by working in the Heisenberg picture.  The Heisenberg evolved
operators depend only on the unitary dynamics of the system, i.e.\ the
protection Hamiltonian and the measurement interaction, and are thus
independent of the state of the system.

Consider a system with canonical operators $(\hat q', \hat p')$ and
suppose that Bob wishes to protect the ``coherent state''
\begin{equation}
  \label{hamstate}
  \ket{\psi_{c_q,c_p}} = \frac1{\pi^{1/4}} \int_{-\infty}^\infty
  \exp\left(-\frac{(q'-c_q)^2}{2} + iq'c_p\right)\ket{q'}dq', 
\end{equation}
where $c_q$ and $c_p$ are constants.  Note that these states are
nonorthogonal for different values of $c_q$ and $c_p$, so this class
of states is sufficient to see that Hamiltonian protected measurements
can distinguish nonorthogonal states.

Bob's protects the system by setting its Hamiltonian to be a (displaced) Harmonic oscillator:
\begin{equation}
  \label{systemham}
  \hat H_S = \frac12\left((\hat p' - c_p)^2 + (\hat q' -
    c_q)^2\right). 
\end{equation}
so that $\ket{\psi_{c_q,c_p}}$ is its nondegenerate ground state with
eigenvalue $1/2$.

Alice wishes to measure the quadrature observable
$\hat a_{\theta} = \cos \theta \hat q' + \sin \theta \hat p'$, which
has expectation value $c_{\theta} = c_q \cos \theta + c_p \sin \theta$
in the state $\ket{\psi_{c_q,c_p}}$.  She does this by coupling the
system to a pointer, with canonical operators $(\hat Q, \hat P)$, via
the interaction Hamiltonian $H_I = g \hat a_{\theta} \otimes \hat P$
for a time duration $1/g$.

Let us change the system co-ordinates to
$\hat q = (\hat q' - c_q) \cos \theta + (\hat p' - c_p) \sin \theta$
and
$\hat p = - (\hat q' - c_q) \sin \theta + (\hat p' - c_p) \cos
\theta$,
noting that $[\hat q, \hat p] = [\hat q', \hat p'] = i$. In these
co-ordinates, the overall Hamiltonian is now
\begin{equation}
  \hat H  = \frac12\left(\hat p^2 + \hat q^2\right) +
  g (\hat q + c_\theta) \otimes \hat P. 
\end{equation}

Heisenberg's equation gives
\begin{equation}
  \frac{d}{dt}\begin{pmatrix}\hat q \\ \hat p \\ \hat Q \\ \hat P\end{pmatrix}
  =
  i\begin{pmatrix}[\hat H, \hat q] \\ [\hat H, \hat p] \\ [\hat H,
    \hat Q] \\ [\hat H, \hat P]\end{pmatrix}
  =
    \begin{pmatrix}\hat p \\ -\hat q - g \hat P \\ g (\hat q + c_\theta) \\ 0\end{pmatrix},
  \label{heisenberg}
\end{equation}
with solution:
\begin{equation}
  \begin{pmatrix}\hat q(t) \\ \hat p(t) \\ \hat Q(t) \\ \hat P(t)\end{pmatrix}
  = 
  \begin{pmatrix}
    \hat q_t + g (\cos t-1)\hat P(0) \\
    \hat p_t - g (\sin t)\hat P(0) \\
    \hat Q(0) + g \left(c_\theta t + \hat p(0) - \hat p_t\right) + g^2
    (\sin t - t)\hat P(0) \\
    \hat P(0)
  \end{pmatrix},
\end{equation}
where we have used the notation $\hat q_t = \hat q(0)\cos t + \hat
p(0)\sin t$ and $\hat p_t = -\hat q(0)\sin t + \hat p(0)\cos t$
based on the solution without the measurement interaction.

When the measurement is complete at $t=1/g$, the solution reads:
\begin{equation}
\begin{pmatrix}\hat q(1/g) \\ \hat p(1/g) \\ \hat Q(1/g) \\ \hat P(1/g)\end{pmatrix}
= 
  \begin{pmatrix}
    \hat q_{1/g} + g(\cos (1/g) - 1)\hat P(0) \\
    \hat p_{1/g} - g(\sin (1/g))\hat P(0) \\
    \hat Q(0) + c_\theta + g \left(\hat p(0) - \hat p_{1/g} - \hat
      P(0)\right) + g^2 (\sin (1/g))\hat P(0) \\
    \hat P(0)
  \end{pmatrix},
\end{equation}
in particular when
\begin{enumerate}
\item\label{von} $g \rightarrow \infty$:
  \begin{equation}
    \begin{pmatrix}\hat q(1/g) \\ \hat p(1/g) \\ \hat Q(1/g) \\ \hat P(1/g)\end{pmatrix}
    = 
    \begin{pmatrix}
      \hat q(0) \\
      \hat p(0) - \hat P(0) \\
      \hat Q(0) + \hat{q}(0) + c_\theta  \\
      \hat P(0)
    \end{pmatrix}.
  \end{equation}
\item\label{semiprotected} $g = 1/(2\pi n)$ for $n = 1,2,\ldots$:
  \begin{equation}
    \begin{pmatrix}\hat q(1/g) \\ \hat p(1/g) \\ \hat Q(1/g) \\ \hat P(1/g)\end{pmatrix}
    = 
    \begin{pmatrix}
      \hat q(0) \\
      \hat p(0) \\
      \hat Q(0) + c_\theta - g\hat P(0) \\
      \hat P(0)
    \end{pmatrix},
  \end{equation}
\item\label{protected} $g \rightarrow 0$:
  \begin{equation}
    \begin{pmatrix}\hat q(1/g) \\ \hat p(1/g) \\ \hat Q(1/g) \\ \hat P(1/g)\end{pmatrix}
    = 
    \begin{pmatrix}
      \hat q_{1/g} \\
      \hat p_{1/g} \\
      \hat Q(0) + c_\theta\\
      \hat P(0)
    \end{pmatrix}.
  \end{equation}
\end{enumerate}

\Cref{von} is easily recognised as the standard von Neumann scheme for
measuring $\hat a_\theta = \hat q + c_\theta$. A measurement of
$\hat Q(1/g)$ behaves like a measurement of $\hat a_\theta$ with an
error given by $\hat Q(0)$. If we prepare the pointer in a state
sharply peaked at $Q=0$ then the measurement will be very accurate
but, by the uncertainty principle, there will be a large spread in
$\hat P$, and hence a large disturbance to the system variable
$\hat p$---this is the usual projective or strong measurement. If we
prepare the ancilla in a state sharply peaked at $P=0$ then
disturbance to $\hat p$ will be small but by the uncertainty principle
there will be a large spread in $\hat Q$ and hence the measurement
will be inaccurate; this is a ``weak'' measurement.

\Cref{protected} is the standard Hamiltonian protected measurement
scheme. Notice that the system undergoes only its free evolution under
$\hat H_S$. Hence, if the quantum state is an eigenstate of
$\hat H_S$, its quantum state is unchanged. A measurement of
$\hat Q(1/g)$ indeed gives access to $c_\theta$ (with error based on
the spread in $\hat Q(0)$), but this is just a parameter in the
Hamiltonian.  This happens regardless of the initial state of the
system, so we can say quite definitively that the information about
$c_{\theta}$ is coming entirely from the protection operation rather
than from the system itself \footnote{Formally, if we specify an
  initial state of the pointer and then cast a final measurement of
  $\hat Q$ as a POVM on the system, all of the POVM will be
  proportional to the identity.}.

Note that we have only proved that this happens for protective
measurements of quadrature observables for the class of states given
in \cref{hamstate} and the corresponding protection Hamiltonians given
in \cref{systemham}.  However, quadrature measurements suffice to
completely determine the state of the system, and these states are
non-orthogonal, so they would not be perfectly distinguishable by
ordinary quantum measurements.  If there is an argument that
Hamiltonian protected measurements imply the reality of the quantum
state in general then it ought to apply to this setup in particular,
but in this case it is clear that all the information comes from the
protection Hamiltonian.

Finally, from \cref{semiprotected} we see that for a carefully chosen,
but nonzero, interaction strength, e.g.\ $g = 1/(2\pi)$, which
corresponds to a finite time duration measurement, the system is
undisturbed (for \emph{any} initial state). If we prepare the pointer
in a state sharply peaked at $Q - gP = 0$ then we can learn $c_\theta$
with arbitrary accuracy.  This shows that criticisms of Hamiltonian
protected measurement based on analysis of practical considerations
are misdirected.  The criticism is that Hamiltonian protected
measurement only works exactly in the limit $g \rightarrow 0$, in
which case the measurement would take an infinite amount of time.  For
finite duration procedures, the state of the system becomes entangled
with the measurement device and is hence disturbed \cite{schcla,
  hagar}.  To perform a practical protective measurement, Alice would
need to know how small to set $g$ in order to make this disturbance
negligible, and it has been argued that this requires knowledge of the
Hamiltonian that is tantamount to knowing $\Ket{\psi}$ in the first
place \cite{hagar}.

However, we have now seen that, for a particular class of
nonorthogonal states and protection Hamiltonians, the same effect as
an exact protective measurement can be achieved in a fixed finite time
duration that is independent of the state and Hamiltonian.  Thus, if
there is an argument that protective measurements do not imply the
reality of the quantum state, it cannot be that finite duration
protective measurements necessarily disturb the quantum state.

\section{Toy models}

\label{Toy}

Our arguments so far have shown that most of the information about the
unknown state of the system (all of it in the Hamiltonian case) comes
from the protection operation rather than the system.  This should be
enough to convince most people that protective measurement cannot have
any implications for the reality of the quantum state.  However, we
have yet to demonstrate formally that protective measurements can be
achieved within models that are $\psi$-epistemic according to the
definitions used in the recent $\psi$-ontology theorems.

To achieve this, we construct $\psi$-epistemic toy models of
protective measurements in theories that have a well-defined state of
reality (ontic state), otherwise known as \emph{ontological}
models\footnote{You may alternatively call these ``classical'' models
  or ``hidden variable theories'', depending on your personal
  terminology preferences.}.  In this framework, quantum states are
represented by probability measures over the ontic states.  The model
is called $\psi$-ontic if the probability measures corresponding to
non-orthogonal states do not overlap and is otherwise $\psi$-epistemic
(see \cite{nic,psireview} for further discussion of this definition).

For Zeno protected measurements, we do not try to exactly reproduce
the quantum predictions, but give a toy model for a qubit that
reproduces the salient features.  Namely, non-orthogonal quantum
states are represented by overlapping probability measures, there is a
protection operation corresponding to repeated projective
measurements, and it is possible to measure the expectation values of
enough observables to determine the quantum state without disturbing
the system by coupling to a continuous variable pointer system.  This
model is constructed by modifying Spekkens' well known
$\psi$-epistemic toy theory \cite{toytheory} to allow for continuous
coupling to a pointer system.

For Hamiltonian protected measurements, we exhibit a toy theory that
reproduces the predictions of the example used in \S\ref{Exact:Ham}
exactly.  This just exploits the fact that all the states and
Hamiltonians involved are Gaussian, and it is known that Gaussian
quantum mechanics can be reproduced by a $\psi$-epistemic model
\cite{erl}.  Whilst this is not a new theory, it is instructive to
track exactly what happens to the ontic state of a system during the
course of a protective measurement in this theory.  We note that weak
values have been analysed by Karanjai \emph{et al} \cite{Karanjai} in
the Gaussian toy theory in a similar way.

The main lesson of both of these models is that protection can be
thought of as an operation that effectively re-prepares the system in
its initial state.  Thus, determining the quantum state of a system by
protective measurements is in closer analogy to performing state
tomography on multiple systems that are independently prepared in the
same state\footnote{See \cite{statedefinetti} for an interpretation of quantum state tomography compatible with the $\psi$-epistemic position.} than it is to having just a single copy of the system.

\subsection{Zeno protected measurements}

\label{Toy:Zeno}

We will start by constructing a toy-model for state preparations and
measurements of a spin-$1/2$ particle in the $x$ and $y$
directions\footnote{We could easily include $z$-measurements as well,
  but having two nonorthogonal states is sufficient for determining
  whether protective measurement entails the reality of the quantum
  state.}.  This model is based on the Spekkens' toy bit \cite{toytheory},
with some modifications to allow for the continuous coupling needed
for protective measurements.

Consider a system with an ontic state space consisting of two random
variables, $X$ and $Y$, that each take values $\pm 1$.  We denote the
state where $X=x$ and $Y=y$ as $(x,y)$ and use $\pm$ as shorthand for
$\pm 1$, so the four possible ontic states are $(+,+)$, $(+,-)$,
$(-,+)$, and $(-,-)$.

\begin{figure}[htb]
  \centering
  \includegraphics[width=0.25\textwidth]{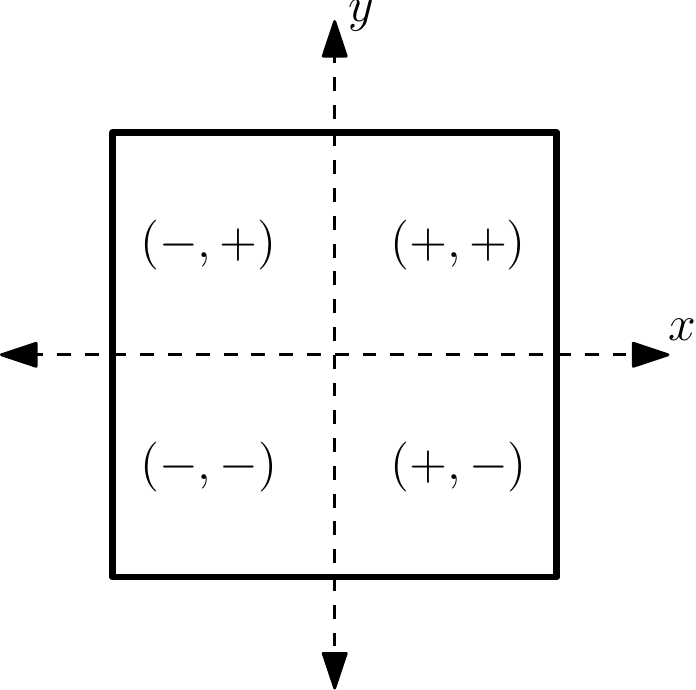}
  \caption{\label{box}The ontic state of the system can be thought of
    as a ball that can be in one of four quadrants of a box.  Here, we
    are looking down on the box along the $z$-axis and the solid line
    indicates the border of the box.}
\end{figure}

For concreteness, we can imagine that the system consists of a ball in
an opaque box with equal width and breadth, where we place the origin
of the $x$-$y$ coordinate system at the center of the box, as
illustrated in Fig.~\ref{box}.  The four possible ontic states then
represent which quadrant of the $x$-$y$ coordinate system the ball is
in, e.g.\ $(+,-)$ represents the state of affairs in which the ball is
in the lower right quadrant, with positive $x$-coordinate and negative
$y$-coordinate.

\begin{figure}[htb]
  \centering
  \includegraphics[width=0.25\textwidth]{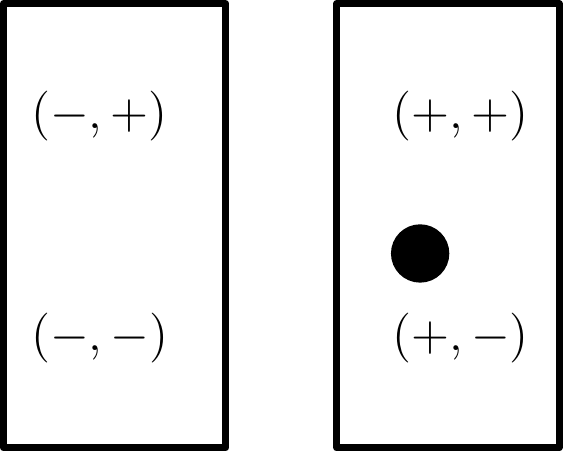}
  \caption{\label{xmeas}An $X$ measurement.  The box can be split in
    two by placing a double partition along the $y$-axis.  The two
    pieces are then shaken vigorously and the sign of the
    $x$-coordinate of the ball can be determined by noting which piece
    of the box rattles.  Doing so causes the $y$-coordinate of the to
    be randomized, so if, as in the diagram, the ball is initially in
    $(+,-)$, it will be in either $(+,-)$ or $(+,+)$ with equal
    probability after the shaking.}
\end{figure}

Now imagine that we do not have complete control over the position of
the ball within the box.  In fact, suppose that there are only two
things we can do to it.  Firstly, we can place a double-partition
along the $y$-axis and separate the box into two pieces according to
the sign of the $x$-coordinate (see Fig.~\ref{xmeas}).  We can then
take each of the two pieces and shake them vigorously.  By noting
which of the pieces rattles when we shake it, we can determine whether
the $x$-coordinate is positive or negative, but doing so causes the
$y$-coordinate to be randomized from the shaking.  We call this an $X$
measurement.  Alternatively, we can place a double partition along the
$x$-axis, separate the box according to the sign of the
$y$-coordinate, and do the same thing.  This allows us to determine
the sign of the $y$-coordinate at the expense of randomizing the
$x$-coordinate.  We call this a $Y$ measurement.

In general, we can describe our knowledge about the ontic state of the
system at any given time via a probability distribution $\bm{p} =
(p_{++},p_{+-},p_{-+},p_{--})$, where $p_{xy}$ denotes the probability
that the ball is in state $(x,y)$.  If, at the start, we know nothing
about where the ball is then the probability distribution will be
$\bm{p} = (1/4,1/4,1/4,1/4)$.  By performing an $X$ measurement and
postselecting on the cases where the $x$-coordinate is found to be
positive, we can prepare the system in the state $\bm{p}^{x+} =
(1/2,1/2,0,0)$, and by postselecting on the cases where it is found to
be negative we can prepare $\bm{p}^{x-} = (0,0,1/2,1/2)$.  Similarly,
with $Y$ measurements we can prepare the states $\bm{p}^{y_+} =
(1/2,0,1/2,0)$ and $\bm{p}^{y-} = (0,1/2,0,1/2)$.

It is easy to see that preparing the system in one of these states
followed by performing a sequence of $X$ and $Y$ measurements has the
same statistics as preparing a spin-$1/2$ particle in the states
$\Ket{x\pm}$ or $\Ket{y\pm}$ followed by performing a sequence of
measurements of the spin in the $x$ and $y$ directions.  Therefore, we
can regard this system as an ontological model, or simulation, of such
experiments on a spin-$1/2$ particle.

In this model, the pure states $\Ket{x\pm}$, $\Ket{y\pm}$ are
represented by probability distributions that are spread out over two
of the four possible ontic states.  Further, the $x$ and $y$ states
overlap, e.g.\ $\bm{p}^{x+}$ and $\bm{p}^{y+}$ both assign probability
$1/2$ to the ontic state $(+,+)$.  Therefore, the quantum state is
epistemic in this model.  Given full knowledge of the ontic state of
the system, it is impossible to tell with certainty which quantum
state was prepared.

The expectation values of $X$ and $Y$ for the four states we can
prepare are shown in Table~\ref{expect}.  Note that the expectation
values completely determine the state so if we can find a method,
analogous to protective measurement, of measuring these expectation
values without disturbing the state then would be able to determine
the state with just a single copy of the system.  However, since the
quantum state is epistemic in the model, this would show that this
feature of Zeno protected measurement does not entail the reality of
the quantum state.

\begin{table}[htb]
  \centering
  \begin{tabular}{|c|r|r|}
    \hline
    State & $\langle X \rangle$ & $\langle Y \rangle$ \\
    \hline
    $\bm{p}^{x+}$ & +1 & 0 \\
    $\bm{p}^{x-}$ & -1 & 0 \\
    $\bm{p}^{y+}$ & 0 & +1 \\
    $\bm{p}^{y-}$ & 0 & -1 \\
    \hline
  \end{tabular}
  \caption{\label{expect}Expectation values of the $X$ and $Y$
    measurements for each of the four states we can prepare.}
\end{table}

In our analogue of Zeno protective measurement, Bob's protection comes
from repeated strong measurements; an $X$ measurement to protect
$\bm{p}^{x_{\pm}}$ or a $Y$ measurement to protect $\bm{p}^{y\pm}$.
However, we also need a model for Alice's measurements, i.e.\ we need
to know how to continuously couple a pointer to the observable we want
to measure, such that the amount of disturbance is small if the
interaction strength is small and the interaction only acts for a
short time interval.  We turn to this next.

Since our toy model is essentially a classical system, we can in
principle perform measurements without any back-action.  This type of
model is enough to make our point, but we will later describe a model with
back-action in order to simulate quantum measurements more precisely.

Our model for a classical measuring device will be a classical point
particle in one-dimension, with position $Q$ and momentum $P$.
Suppose that the particle is initially prepared in the state $Q=P=0$
at time $t=0$.  When we want to measure the $x$ coordinate of the
system, we couple the momentum $P$ of the particle to the $X$ variable
with the interaction Hamiltonian $H_I = g XP$ from time $t=0$ to
$t=1/g$, where $g$ is a coupling constant.

With this interaction, the particle will move one unit to the right if
$X=+1$ and one unit to the left if $X=-1$, so long as the ball remains
undisturbed during the course of the measurement, and the pointer will
reach this position at $t = 1/g$.  Thus, if we prepare the system in
the state $\bm{p}^{x\pm}$ the pointer will end up at $Q=\pm 1$ with
certainty, whereas if we prepare the system in either of the states
$\bm{p}^{y\pm}$ then the pointer will move to $Q = +1$ with
probability $1/2$ and $Q = -1$ with probability $1/2$ because, in
those states, the $x$-coordinate of the system is either positive or
negative with $50/50$ probability.

Similarly, to measure the $y$-coordinate of the system, we use the
Hamiltonian $H_I = g YP$ in the same way, and now the pointer will
move to $Q = \pm 1$ with certainty when the $\bm{p}^{y \pm}$ states
are prepared and will move randomly either to $Q=+1$ or $Q=-1$ when
the $\bm{p}^{x \pm}$ states are prepared.

We can now describe an analogue of Zeno protected measurements in our
model.  Suppose that the system is initially prepared in one of the
states $\bm{p}^{x\pm}$ or $\bm{p}^{y\pm}$ and that these are protected
by performing either strong $X$ measurements or strong $Y$
measurements respectively at times $t_n = n\Delta t$ for
$n = 0,1,\ldots,N$, where $\Delta t = 1/gN$.  At the same time, the
system is coupled to the measuring device by one of the continuous
processes described above.  In the limit $N \rightarrow \infty$, we
will show that the pointer ends up pointing to the expectation value
of the quantity being measured with probability $1$, and the system
remains in the probability distribution that it started in, i.e.\
either $\bm{p}^{x\pm}$ or $\bm{p}^{y\pm}$.  If we repeat this process
twice, coupling the pointer to $X$ the first time round and $Y$ the
second time round then we can obtain the expectation values of both
observables, which serve to identify which probability distribution
was prepared with certainty.  Thus, if protective measurement were a
proof of the reality of the quantum state then this model would
analogously be a proof of the reality of the probability distributions
$\bm{p}^{x\pm}$ and $\bm{p}^{y\pm}$.  Since these distributions
overlap, i.e.\ the ontic state is not enough to determine which
distribution was prepared, this is clearly preposterous.  What is
actually happening is that the protective measurement is mostly
measuring a property of the measurements that are protecting the
system, rather than a property of the initial ontic state of the
system.  It is the randomization due to this protection that causes
the pointer to point to the expectation value.

Let's see how this works in a little more detail.  Suppose the system
is prepared in $\bm{p}^{x+}$ and is protected by strongly measuring
$X$ every $\Delta t$ seconds.  This will cause the $y$-coordinate to
be randomized every $\Delta t$ seconds, whilst leaving the
$x$-coordinate as it is.  If the continuous measurement interaction is
set to measure $X$ then the randomization of the $y$-coordinate will
have no effect on the motion of the pointer because the Hamiltonian
$H_I = g XP$ only couples to the $x$-coordinate.  Thus, the
measurement will work just as it did without the protection and the
pointer will move to $Q=+1$ in time $1/g$.  This is the expectation
value of $X$ in this case.

If, on the other hand, the continuous measurement interaction is set
to measure $Y$ then the $y$-coordinate is initially random, so the
pointer will move $1/N$ units to the right or left with $50/50$
probability in the time interval before the first strong $X$
measurement.  After each strong $X$ measurement, the $y$-coordinate is
randomized again, so there will be a probability $1/2$ that the
pointer continues moving in the same direction for another distance
$1/N$ and a probability $1/2$ that it switches direction and moves
$1/N$ units in the opposite direction.  If we denote the position of
the pointer at time $t_n$ as $Q_n$, the variables $Q_n$ describe a
$N$-step discrete time random walk on the line with step distance
$1/N$.

The quantity $\tilde{Q}_n = N(Q_n - Q_{n-1})$ describes the change in
position of the pointer in the $n$th time step, rescaled such that
$\tilde{Q}_n = \pm 1$.  The variables $\tilde{Q}_n$ are independently
and identically distributed with uniform distribution because the
$y$-coordinate is freshly randomized at every step.  We can now write
the final position of the pointer as
\begin{equation}
  Q_N = \frac{1}{N}\sum_{n = 1}^N \tilde{Q}_n,
\end{equation}
from which we see that it is the sample mean of $N$ i.i.d.\ variables
with uniform distribution.  The variables $\tilde{Q}_n$ have mean $0$
and variance $1$ so, by the central limit theorem, in the limit $N
\rightarrow \infty$ the distribution of $Q_N$ approaches a normal
distribution with mean $0$ and variance $1/N$.  Since the variance
tends to zero as $N \rightarrow \infty$, the limit will be a Dirac
$\delta$ distribution centred at $0$.  So, in this limit, the pointer
will end up pointing to $Q=0$ with probability $1$, and this is the
expectation value of $Y$.

Note that, in both of these measurements, the system remains in the
$\bm{p}^{x+}$ distribution throughout.  Therefore, we can perform a
protective measurement of $X$ followed by a protective measurement of
$Y$ and obtain both the expectation values.  Similarly, if we started
with $\bm{p}^{x-}$ or $\bm{p}^{y\pm}$ then we could obtain both $X$
and $Y$ expectation values in the same way.  This would allow us to
tell with certainty which of the four states had been prepared.

Obviously, without the protection, there is no procedure that would
allow $\bm{p}^{x+}$ to be distinguished from $\bm{p}^{y+}$ with
certainty, as both distributions assign probability $1/2$ to the ontic
state $(+,+)$.  Performing a $Y$ measurement on a system prepared in
$\bm{p}^{x+}$ without protection would yield the values $\pm 1$ with
$50/50$ probability rather than the expectation value.  The reason
that we get the expectation value $0$ with protection is that the
protecting measurement randomizes the $y$-coordinate, which
effectively reprepares the system in an independent copy of
$\bm{p}^{x+}$ each time.  Thus, Zeno protective measurement is far
more like distinguishing $N$ copies of $\bm{p}^{x+}$ from $N$ copies
of $\bm{p}^{y+}$ than it is distinguishing a single copy, and this can
be done with with arbitrary precision in the limit
$N \rightarrow \infty$.

Here, we can also uncover an implicit assumption in the argument that
Zeno protective measurement implies the reality of the quantum state.
Namely, since a measurement of the Pauli observable $\hat{X}$ on the
state $\Ket{x+}$ does not change the state, it is implicitly assumed
that this means that no property of the system has changed at all so,
in particular, we still have the same single copy of $\Ket{x+}$ that
we started with.  However, in our model there are two ontic states,
$(+,+)$ and $(+,-)$, that can be occupied when we prepare
$\bm{p}^{x+}$ and the protecting measurement randomly switches them.
Thus, after a protecting measurement, the $y$-coordinate is completely
uncorrelated from what it was before, so the protective measurement
re-prepares the system in a totally independent copy of $\bm{p}^{x+}$.

This concludes our basic model of Zeno protected measurements. However, in the model constructed so far, our continuous measurements do not simulate the disturbance to the system
caused by a quantum measurement.  To simulate quantum theory more
closely, we would like a model in which the coordinate that is not
being measured gets gradually randomized during the course of the
measurement, so that it is completely randomized at time $t = 1/g$.

To model this, suppose that, during the course of a measurement of the
$x$-coordinate, the system is subjected to a continuous Markovian
evolution, such that the probability of making a transition from
$(x,+)$ to $(x,-)$ or vice versa in a time interval $\D t$ is $r\D t$,
where $r$ is an arbitrary transition rate parameter that we may set as
we please.  The probability distribution is then governed by the
master equations
\begin{align}
  \frac{\D p_{x+}}{\D t} & = -rp_{x+} + rp_{x-} & \frac{\D p_{x-}}{\D
    t} = -rp_{x-} + rp_{x+},
\end{align}
which have solution
\begin{align}
  \label{backaction}
  p_{x+}(t) & = \frac{1}{2} \left [ p_{x+}(0) \left ( 1 + e^{-2rt}
    \right ) + p_{x-}(0) \left (1 - e^{-2rt} \right ) \right ] \\
  p_{x-}(t) & = \frac{1}{2} \left [ p_{x-}(0) \left ( 1 + e^{-2rt}
    \right ) + p_{x+}(0) \left (1 - e^{-2rt} \right ) \right ].
\end{align}

In the limit $t \rightarrow \infty$ this gives 
\begin{equation}
  p_{x+} = p_{x-} = \frac{1}{2} [ p_{x+}(0) + p_{x-}(0) ],
\end{equation}
so the $y$-coordinate gets completely randomized, whilst the
$x$-coordinate remains unaffected.  If we imagine that this process is
going on at the same time as the Hamiltonian coupling $H_I = g X P$
then the pointer will move just as it did before because $X$ does not
change, but in the limit $g \rightarrow 0$, the back-action will have
enough time to completely randomize the $y$-coordinate.  Thus, by
combining the Hamiltonian coupling with this back-action, we get the
same effect as performing an instantaneous $X$ measurement by shaking
the two parts of the box.

Incorporating this disturbance has surprisingly little effect on the analysis of protective measurements. Consider again a system prepared in $\bm{p}^{x+}$, protected with
strong $X$ measurements, and continuously coupled to a pointer
measuring $X$.  In this case, the back action causes the system to
switch between $(+,+)$ and $(+,-)$.  This does not affect the motion
of the pointer, which still moves continuously towards $Q=+1$, because
the pointer is only coupled to the $x$-coordinate.  It does not affect
how the probability distribution of the system evolves because
distributions with $p_{++} = p_{+-}$ are stationary states of the
back-action evolution given in Eq.~\eqref{backaction}.

The more interesting case is where we continuously measure $Y$ on a
protected system prepared in $\bm{p}_{x+}$.  Again, the back-action
does not affect the motion of the pointer, as this is coupled to $Y$
and the back action of a continuous $Y$ measurement only affects the
$x$-coordinate.  On the other hand, for a finite $N$, the back action
causes the protecting measurement to sometimes fail, just as it does
in the quantum case.  This is because it is now possible for a
transition of the $x$-coordinate to occur, and when this happens the
protecting measurement yields $X=-1$.  However, the failure
probability can be made arbitrarily small by taking the limit
$N \rightarrow \infty$, just as in the quantum case.

In a bit more detail, under the $Y$-measurement back-action, the
probability that the system remains in an ontic state with positive
$x$-coordinate after time $\Delta t$ is
\begin{align}
  p_{X=+1}(\Delta t) & = p_{++}(\Delta t) + p_{+-}(\Delta t) \\
  & = \frac{1}{2} \left [ p_{++}(0)(1+e^{-2 r\Delta t}) + p_{-+}(0)
    (1-e^{-2r\Delta t}) + p_{+-}(0)(1+e^{-2 r \Delta t}) + p_{--}(0)
    (1 - e^{-2r \Delta t})\right ].
\end{align}
For a starting state of $\bm{p}^{x+}$ we have $p_{++}(0) = p_{+-}(0) =
1/2$ and $p_{-+}(0) = p_{--}(0) = 0$, so
\begin{equation}
  p_{X=+1}(\Delta t) = \frac{1}{2} \left ( 1 + e^{-2 r \Delta t}\right ).
\end{equation}
If there are a total of $N$ protecting measurements during the course
of the measurement then the probability that every protecting
measurement gives the $X = +1$ outcome is $p_{X=+1}(\Delta t)^N$.
This gives a probability of success of
\begin{equation}
  p_{\text{succ}} = \left [ \frac{1}{2} \left ( 1 + e^{-2r/g N} \right )\right ]^N.
\end{equation}
We still have the freedom to set $r$ as a function of $g$ and $N$, and
setting $r = g / 2 N^3$ gives a success probability that converges to
$1$ as $N \rightarrow \infty$.

\subsection{Hamiltonian protected measurements}

\label{Toy:Ham}

A natural $\psi$-epistemic ontological model for the subset
of quantum theory used in \cref{Exact:Ham} has already been
proposed \cite{erl}. In short, for this subset (``Gaussian quantum
mechanics'') the Wigner representation of states, transformations, and
measurements are all non-negative and therefore admit a probabilistic
interpretation in terms of a classical phase space, in this case $(q',
p', Q, P)$.

The states $\ket{\psi_{c_q,c_p}}$ correspond to Gaussian probability
distributions
\begin{equation}
  P_{c_q,c_p}(q',p') = \frac{1}{\pi}\exp\left(-(q'-c_q)^2 - (p' - c_p)^2\right).
\end{equation}

\begin{figure}[htb]
\includegraphics[scale=0.75]{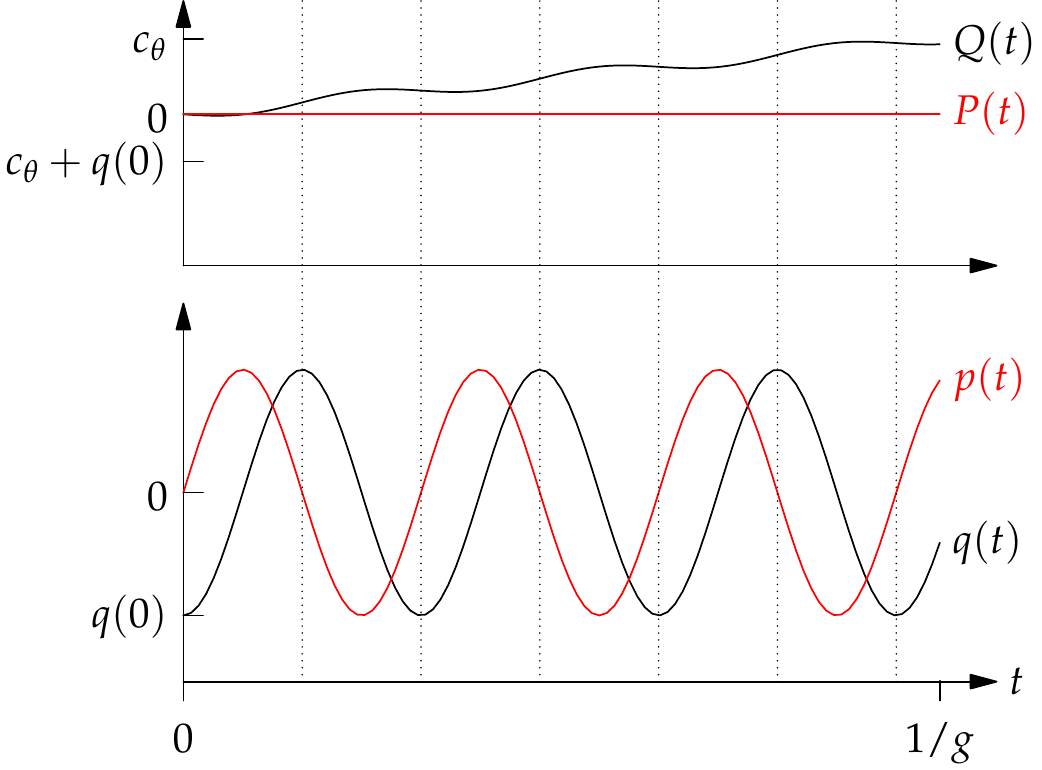}
\qquad
\includegraphics[scale=0.75]{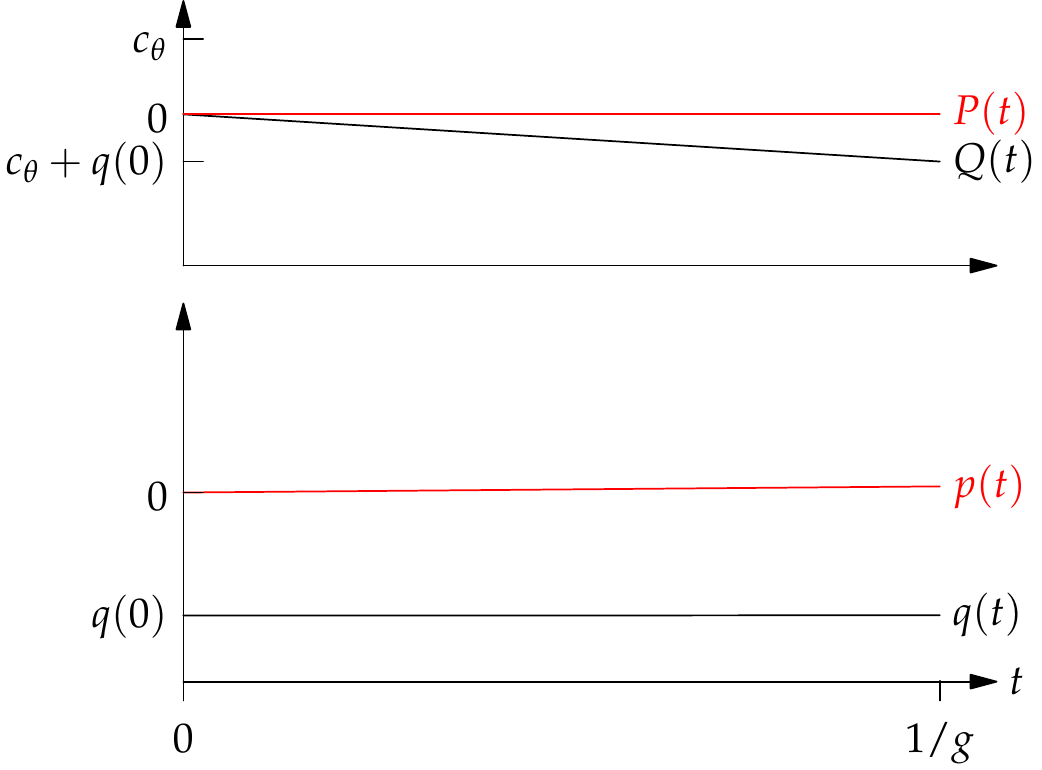}
\caption{\label{toyplot}What happens during measurements of $\hat a' = c_\theta + \hat q$ according to $\psi$-epistemic model, in the simple case $P=0$ where there is no disturbance to the system due to the measurement interaction. On the left is a protective measurement, which due to the time-averaging of the system's free evolution gives the parameter $c_\theta$ regardless of the system's initial state. On the right is a traditional von-Neumann type measurement which gives the actual value of $c_\theta + q$.}
\end{figure}

\begin{figure}[htb]
\includegraphics[scale=0.75]{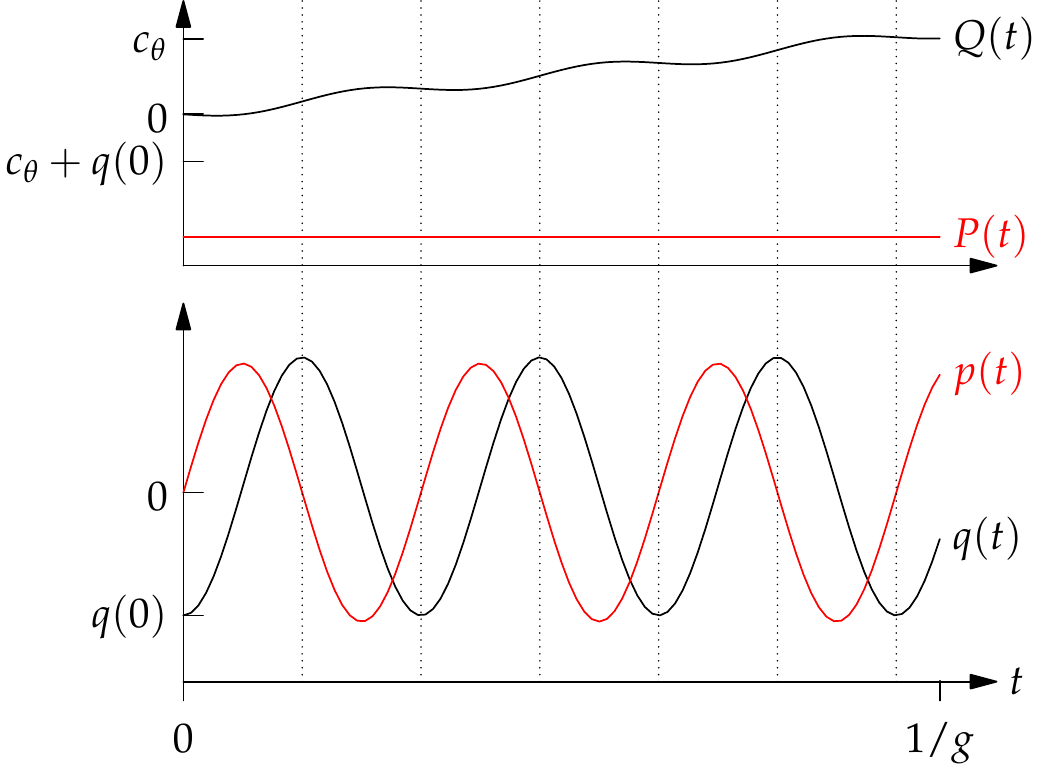}
\qquad
\includegraphics[scale=0.75]{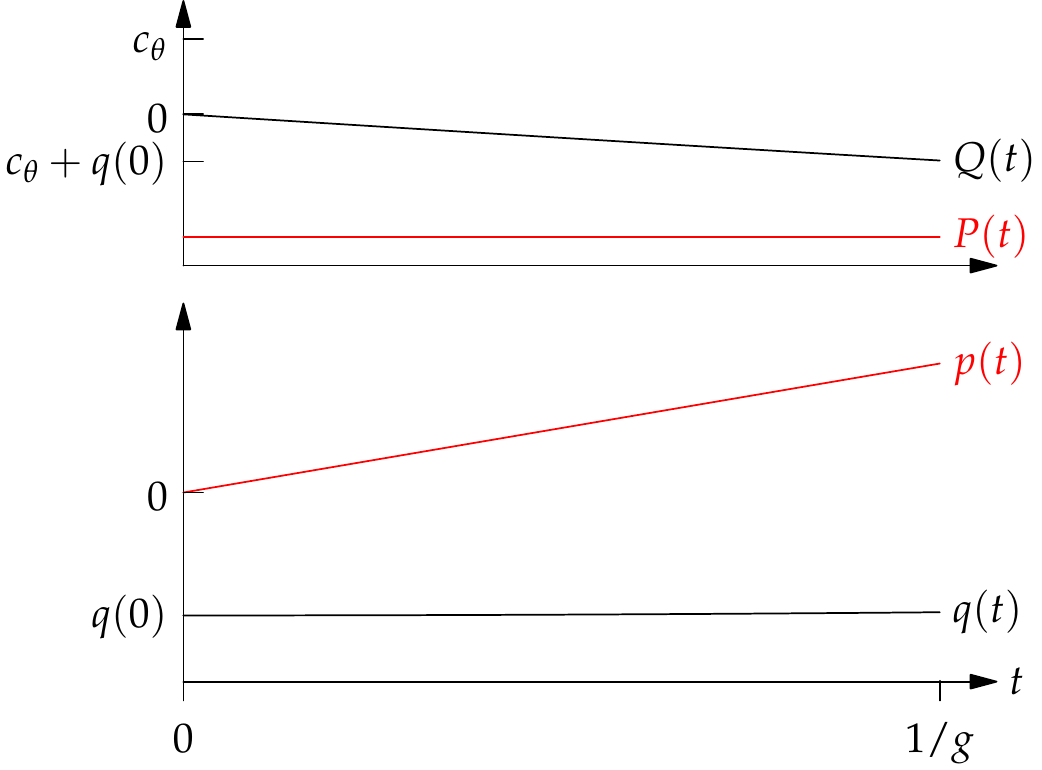}
\caption{\label{toyplot2}As in \cref{toyplot}, but now with non-zero $P(0)$, so that $p$ is disturbed by the measurement. In the protective case on the left, the free evolution ``smears out'' the disturbance.}
\end{figure}

The time evolution follows the classical Hamilton's equations, which
matches the evolution of the operators in \eqref{heisenberg} (with the
new phase space variables $(q,p,Q,P)$ defined in an analogous way to
the corresponding operators)
\begin{equation}
  \frac{d}{dt}\begin{pmatrix}q \\ p \\ Q \\ P\end{pmatrix}
    =
    \begin{pmatrix}\{q, \mathcal{H}\} \\ \{p, \mathcal{H}\} \\ \{Q, \mathcal{H}\} \\ \{P, \mathcal{H}\}\end{pmatrix}
    =
    \begin{pmatrix}p \\ -q - gP \\ g(q + c_\theta) \\ 0\end{pmatrix}.
\end{equation}
Solutions to these equations are illustrated in Figs.~\ref{toyplot}
and \ref{toyplot2}.

The story is simplest when $P=0$. We see that the derivative of $Q$ is
proportional to $a_\theta$, the variable we are protectively
measuring. But meanwhile the system variables $(q,p)$ are evolving
according to the free evolution, so that $a_\theta$ oscillates around
$c_\theta$. The final pointer position $Q$ is the time-average, which
will be exactly $c_\theta$ in the ``protective measuremnt'' limit of
small $g$ (or equivalently large final time $1/g$). When $P \neq 0$
the system variable $p$ is disturbed by the measuring process, and the
``protecting'' free evolution smears this disturbance, ironically
making the disturbance affect the variable $a_\theta$ we are trying
to measure, but $a_\theta$ still averages to $c_\theta$.


Notice that if we could violate the uncertainty principle by preparing
a pointer system with $Q=0$ and $P=0$, we could measure $a_\theta$
without causing any disturbance to the system \footnote{That is,
  without any disturbance to $(q,p)$. Of course our \emph{state of
    knowledge} about the system would change, but nobody trying to
  learn about a system should want ''protection'' from ''disturbance''
  to their knowledge of the system!}. But without the ``protection''
from the free evolution, we would simply get a sample from
$P_{c_q,c_p}$---which would not uniquely fix $(c_q,c_p)$. In this
model we can see explicitly that the main role of the ``protector'' is
actually to provide a ``time ensemble'' of system states $(q',p')$
whose time averages are $(c_q, c_p)$.

In fact the basic idea of the above was anticipated in the original
paper \cite{meaning}, and the authors offered two responses:
\begin{enumerate}
\item[(a)] If there are nodes in, say, a 1-dimensional position
  wavefunction, the particle would have to travel at infinite speed in
  order to ensure that it is never found there. (This does not arise
  above because Gaussian wavefunctions do not have nodes.).
\item[(b)] Systems in eigenstates of real Hamiltonains have constant
  position in Bohmian mechanics. (This does not arise above because
  the ontological model we use behaves very differently to Bohmian
  mechanics as applied to the Gaussian quantum mechanics.)
\end{enumerate}
The latter can be easily dismissed in the present context - Bohmian
mechanics is a $\psi$-ontic ontological model and so is clearly
useless as a counter-example to claims that protective measurement
establishes the reality of the quantum state. The former is a little
more compelling, but it is unclear how to actually turn this into an
argument for the reality of the wavefunction that is not vunerable to
our counter-example. Indeed such an argument would presumably be
unable to establish the reality of Gaussian wavefunctions (perhaps it
would only establish the reality of wavefunctions with nodes), which
would be a rather odd situation. Nodes in a wavefunction might just be
a quantum phenomena that is difficult to account for in any
ontological model, indepedently of whether it is $\psi$-ontic or
$\psi$-epistemic.

\section{Conclusion}

\label{Conc}

In this paper, we have given three arguments that protective
measurements do not imply the reality of the quantum state.  Firstly,
we pointed out that, given the same resources as in protective
measurement, Alice could determine the state of the system in a much
more straightforward way by doing process tomography on the protection
operation.  Secondly, we showed that most of the information in a
protective measurement comes from the protection operation rather than
the system itself --- in fact all of it for Hamiltonian protected
measurements.  In the course of doing this, we found a new procedure
equivalent to protective measurement that works for a class of
Gaussian states and quadratic Hamiltonians that has a fixed finite
interaction strength and time duration.  This shows that criticisms of
protective measurement based on practical considerations are
misdirected.  Finally, we constructed explicit $\psi$-epistemic
ontological models of protective measurement, which rigorously
establishes that they are compatible with epistemic quantum states.

An implicit underlying assumption of the argument for the reality of
the quantum state based on protective measurement is that anything
which does not change the quantum state does not affect the system at
all.  This idea is already incompatible with a $\psi$-epistemic
interpretation of the quantum state, in which we should think of
quantum states as more akin to probability measures than to classical
fields.  Thus, we expect there to be underlying microstates which may
be disturbed even if the quantum state stays the same.  In fact, this
is the mechanism behind protective measurement in our toy models, in
which the protection operation effectively prepares the system in an
independent copy of the initial state. Indeed, the desire for
``protection'' is itself suspect in the $\psi$-epistemic view, since
it seems based on the converse assumption that anything which changes
the quantum state affects the system. But probability measures can be
updated without any change to the system they describe, and indeed in
our toy models the ``protection'' prevents the measurement from
actually revealing new information about the initial configuration of
the system. If even the nomenclature of protective measurement only
makes sense when presupposing the reality of the quantum state,
proponents of such reality must be especially careful to avoid begging
the question when invoking the protective measurement scheme.

Finally, it would be interesting to extend the analyses of
\S\ref{Exact} and the toy models of \S\ref{Toy} beyond the special
cases considered to arbitrary protective measurements.  In particular,
it would be interesting to determine if the finite duration
Hamiltonian protected measurements of \S\ref{Exact:Ham} exist for more
general classes of states and observables.  Although the present
analysis is more than enough to establish that the naive argument from
``the quantum state can be measured by protective measurements'' to
``the quantum state must be real'' is incorrect, and that analogous
phenomena exist classically, there could be special features of the
particular way that protective measurements work in quantum theory
that depend on genuinely quantum phenomena.  Something similar was
recently shown for the phenomenon of ``anomalous weak values'', which
do have a classical analogue \cite{FerrieCombes14}, but, nonetheless,
the specific way they arise in quantum theory is different to the way
they arise in classical theories \cite{Ipsen15}. In particular
anomolous weak values provide statistical evidence for (or a ``proof''
of) contextuality \cite{Pusey14}.  Thus, we do not wish to claim that
there is nothing ``quantum'' about protective measurements, but rather
that there is no convincing argument that they imply the reality of
the quantum state, and plenty of compelling evidence that they do not.

\begin{acknowledgements}
  MP is grateful to Aharon Brodutch and Shan Gao for discussions, in particular to Shan for correcting MP's initial misunderstanding of the Zeno scheme. Research at Perimeter Institute is supported in part by the
  Government of Canada through NSERC and by the Province of Ontario
  through MRI.  CF was supported by NSF Grant No. PHY-1212445, the Canadian Government through the NSERC PDF program, the IARPA MQCO program, the ARC via EQuS project number CE11001013, and by the US Army Research Office grant numbers W911NF-14-1-0098 and W911NF-14-1-0103. ML is supported by the Foundational Questions
  Institute (FQXi).  We would like to thank Paul Merriam for a careful
  proof reading.
\end{acknowledgements}

\bibliography{protective}

\begin{thebibliography}{58}%
\makeatletter
\providecommand \@ifxundefined [1]{%
 \@ifx{#1\undefined}
}%
\providecommand \@ifnum [1]{%
 \ifnum #1\expandafter \@firstoftwo
 \else \expandafter \@secondoftwo
 \fi
}%
\providecommand \@ifx [1]{%
 \ifx #1\expandafter \@firstoftwo
 \else \expandafter \@secondoftwo
 \fi
}%
\providecommand \natexlab [1]{#1}%
\providecommand \enquote  [1]{``#1''}%
\providecommand \bibnamefont  [1]{#1}%
\providecommand \bibfnamefont [1]{#1}%
\providecommand \citenamefont [1]{#1}%
\providecommand \href@noop [0]{\@secondoftwo}%
\providecommand \href [0]{\begingroup \@sanitize@url \@href}%
\providecommand \@href[1]{\@@startlink{#1}\@@href}%
\providecommand \@@href[1]{\endgroup#1\@@endlink}%
\providecommand \@sanitize@url [0]{\catcode `\\12\catcode `\$12\catcode
  `\&12\catcode `\#12\catcode `\^12\catcode `\_12\catcode `\%12\relax}%
\providecommand \@@startlink[1]{}%
\providecommand \@@endlink[0]{}%
\providecommand \url  [0]{\begingroup\@sanitize@url \@url }%
\providecommand \@url [1]{\endgroup\@href {#1}{\urlprefix }}%
\providecommand \urlprefix  [0]{URL }%
\providecommand \Eprint [0]{\href }%
\providecommand \doibase [0]{http://dx.doi.org/}%
\providecommand \selectlanguage [0]{\@gobble}%
\providecommand \bibinfo  [0]{\@secondoftwo}%
\providecommand \bibfield  [0]{\@secondoftwo}%
\providecommand \translation [1]{[#1]}%
\providecommand \BibitemOpen [0]{}%
\providecommand \bibitemStop [0]{}%
\providecommand \bibitemNoStop [0]{.\EOS\space}%
\providecommand \EOS [0]{\spacefactor3000\relax}%
\providecommand \BibitemShut  [1]{\csname bibitem#1\endcsname}%
\let\auto@bib@innerbib\@empty
\bibitem [{\citenamefont {Bacciagaluppi}\ and\ \citenamefont
  {Valentini}(2009)}]{Bacciagaluppi2009}%
  \BibitemOpen
  \bibfield  {author} {\bibinfo {author} {\bibfnamefont {G.}~\bibnamefont
  {Bacciagaluppi}}\ and\ \bibinfo {author} {\bibfnamefont {A.}~\bibnamefont
  {Valentini}},\ }\href@noop {} {\emph {\bibinfo {title} {{Quantum Theory at
  the Crossroads: Reconsidering the 1927 Solvay Conference}}}}\ (\bibinfo
  {publisher} {Cambridge University Press},\ \bibinfo {year} {2009})\ \Eprint
  {http://arxiv.org/abs/quant-ph/0609184} {arXiv:quant-ph/0609184} \BibitemShut
  {NoStop}%
\bibitem [{\citenamefont {Faye}(2008)}]{Faye2008}%
  \BibitemOpen
  \bibfield  {author} {\bibinfo {author} {\bibfnamefont {J.}~\bibnamefont
  {Faye}},\ }in\ \href
  {http://plato.stanford.edu/archives/fall2008/entries/qm-copenhagen} {\emph
  {\bibinfo {booktitle} {The Stanford Encyclopedia of Philosophy}}},\ \bibinfo
  {editor} {edited by\ \bibinfo {editor} {\bibfnamefont {E.~N.}\ \bibnamefont
  {Zalta}}}\ (\bibinfo {year} {2008})\ \bibinfo {edition} {fall 2008}\
  ed.\BibitemShut {Stop}%
\bibitem [{\citenamefont {Harrigan}\ and\ \citenamefont
  {Spekkens}(2010)}]{nic}%
  \BibitemOpen
  \bibfield  {author} {\bibinfo {author} {\bibfnamefont {N.}~\bibnamefont
  {Harrigan}}\ and\ \bibinfo {author} {\bibfnamefont {R.~W.}\ \bibnamefont
  {Spekkens}},\ }\href {\doibase 10.1007/s10701-009-9347-0} {\bibfield
  {journal} {\bibinfo  {journal} {Found. Phys.}\ }\textbf {\bibinfo {volume}
  {40}},\ \bibinfo {pages} {125} (\bibinfo {year} {2010})},\ \Eprint
  {http://arxiv.org/abs/0706.2661} {arXiv:0706.2661} \BibitemShut {NoStop}%
\bibitem [{\citenamefont {Everett}(1957)}]{Everett1957}%
  \BibitemOpen
  \bibfield  {author} {\bibinfo {author} {\bibfnamefont {H.}~\bibnamefont
  {Everett}},\ }\href {\doibase 10.1103/RevModPhys.29.454} {\bibfield
  {journal} {\bibinfo  {journal} {Rev. Mod. Phys.}\ }\textbf {\bibinfo {volume}
  {29}},\ \bibinfo {pages} {454} (\bibinfo {year} {1957})}\BibitemShut
  {NoStop}%
\bibitem [{\citenamefont {DeWitt}\ and\ \citenamefont
  {Graham}(1973)}]{DeWitt1973}%
  \BibitemOpen
  \bibinfo {editor} {\bibfnamefont {B.~S.}\ \bibnamefont {DeWitt}}\ and\
  \bibinfo {editor} {\bibfnamefont {R.~N.}\ \bibnamefont {Graham}},\ eds.,\
  \href@noop {} {\emph {\bibinfo {title} {{The Many-Worlds Interpretation of
  Quantum Mechanics}}}}\ (\bibinfo  {publisher} {Princeton University Press},\
  \bibinfo {year} {1973})\BibitemShut {NoStop}%
\bibitem [{\citenamefont {Wallace}(2012)}]{Wallace2012}%
  \BibitemOpen
  \bibfield  {author} {\bibinfo {author} {\bibfnamefont {D.}~\bibnamefont
  {Wallace}},\ }\href@noop {} {\emph {\bibinfo {title} {{The Emergent
  Multiverse: Quantum Theory according to the Everett Interpretation}}}}\
  (\bibinfo  {publisher} {Oxford University Press},\ \bibinfo {year}
  {2012})\BibitemShut {NoStop}%
\bibitem [{\citenamefont {de~Broglie}(2009)}]{Broglie2009}%
  \BibitemOpen
  \bibfield  {author} {\bibinfo {author} {\bibfnamefont {L.}~\bibnamefont
  {de~Broglie}},\ }in\ \href@noop {} {\emph {\bibinfo {booktitle} {{Quantum
  Theory at the Crossroads: Reconsidering the 1927 Solvay Conference}}}},\
  \bibinfo {editor} {edited by\ \bibinfo {editor} {\bibfnamefont
  {G.}~\bibnamefont {Bacciagaluppi}}\ and\ \bibinfo {editor} {\bibfnamefont
  {A.}~\bibnamefont {Valentini}}}\ (\bibinfo  {publisher} {Cambridge University
  Press},\ \bibinfo {year} {2009})\ pp.\ \bibinfo {pages} {373--406},\ \bibinfo
  {note} {(page nos. refer to arXiv version)},\ \Eprint
  {http://arxiv.org/abs/quant-ph/0609184} {arXiv:quant-ph/0609184} \BibitemShut
  {NoStop}%
\bibitem [{\citenamefont {Bohm}(1952{\natexlab{a}})}]{Bohm1952}%
  \BibitemOpen
  \bibfield  {author} {\bibinfo {author} {\bibfnamefont {D.}~\bibnamefont
  {Bohm}},\ }\href {\doibase 10.1103/PhysRev.85.166} {\bibfield  {journal}
  {\bibinfo  {journal} {Phys. Rev.}\ }\textbf {\bibinfo {volume} {85}},\
  \bibinfo {pages} {166} (\bibinfo {year} {1952}{\natexlab{a}})}\BibitemShut
  {NoStop}%
\bibitem [{\citenamefont {Bohm}(1952{\natexlab{b}})}]{Bohm1952a}%
  \BibitemOpen
  \bibfield  {author} {\bibinfo {author} {\bibfnamefont {D.}~\bibnamefont
  {Bohm}},\ }\href {\doibase 10.1103/PhysRev.85.180} {\bibfield  {journal}
  {\bibinfo  {journal} {Phys. Rev.}\ }\textbf {\bibinfo {volume} {85}},\
  \bibinfo {pages} {180} (\bibinfo {year} {1952}{\natexlab{b}})}\BibitemShut
  {NoStop}%
\bibitem [{\citenamefont {D{\"u}rr}\ and\ \citenamefont
  {Teufel}(2009)}]{Duerr2009}%
  \BibitemOpen
  \bibfield  {author} {\bibinfo {author} {\bibfnamefont {D.}~\bibnamefont
  {D{\"u}rr}}\ and\ \bibinfo {author} {\bibfnamefont {S.}~\bibnamefont
  {Teufel}},\ }\href
  {http://www.springer.com/physics/quantum+physics/book/978-3-540-89343-1}
  {\emph {\bibinfo {title} {Bohmian Mechanics}}}\ (\bibinfo  {publisher}
  {Springer},\ \bibinfo {year} {2009})\BibitemShut {NoStop}%
\bibitem [{\citenamefont {Ghirardi}\ \emph {et~al.}(1986)\citenamefont
  {Ghirardi}, \citenamefont {Rimini},\ and\ \citenamefont
  {Weber}}]{Ghirardi1986}%
  \BibitemOpen
  \bibfield  {author} {\bibinfo {author} {\bibfnamefont {G.~C.}\ \bibnamefont
  {Ghirardi}}, \bibinfo {author} {\bibfnamefont {A.}~\bibnamefont {Rimini}}, \
  and\ \bibinfo {author} {\bibfnamefont {T.}~\bibnamefont {Weber}},\ }\href
  {\doibase 10.1103/PhysRevD.34.470} {\bibfield  {journal} {\bibinfo  {journal}
  {Phys. Rev. D}\ }\textbf {\bibinfo {volume} {34}},\ \bibinfo {pages} {470}
  (\bibinfo {year} {1986})}\BibitemShut {NoStop}%
\bibitem [{\citenamefont {Bassi}\ \emph {et~al.}(2013)\citenamefont {Bassi},
  \citenamefont {Lochan}, \citenamefont {Satin}, \citenamefont {Singh},\ and\
  \citenamefont {Ulbricht}}]{Bassi2013}%
  \BibitemOpen
  \bibfield  {author} {\bibinfo {author} {\bibfnamefont {A.}~\bibnamefont
  {Bassi}}, \bibinfo {author} {\bibfnamefont {K.}~\bibnamefont {Lochan}},
  \bibinfo {author} {\bibfnamefont {S.}~\bibnamefont {Satin}}, \bibinfo
  {author} {\bibfnamefont {T.~P.}\ \bibnamefont {Singh}}, \ and\ \bibinfo
  {author} {\bibfnamefont {H.}~\bibnamefont {Ulbricht}},\ }\href {\doibase
  10.1103/RevModPhys.85.471} {\bibfield  {journal} {\bibinfo  {journal} {Rev.
  Mod. Phys.}\ }\textbf {\bibinfo {volume} {85}},\ \bibinfo {pages} {471}
  (\bibinfo {year} {2013})},\ \Eprint {http://arxiv.org/abs/1204.4325}
  {arXiv:1204.4325} \BibitemShut {NoStop}%
\bibitem [{\citenamefont {Lombardi}\ and\ \citenamefont
  {Dieks}(2013)}]{Lombardi2013}%
  \BibitemOpen
  \bibfield  {author} {\bibinfo {author} {\bibfnamefont {O.}~\bibnamefont
  {Lombardi}}\ and\ \bibinfo {author} {\bibfnamefont {D.}~\bibnamefont
  {Dieks}},\ }in\ \href
  {http://plato.stanford.edu/archives/fall2013/entries/qm-modal/} {\emph
  {\bibinfo {booktitle} {The Stanford Encyclopedia of Philosophy}}},\ \bibinfo
  {editor} {edited by\ \bibinfo {editor} {\bibfnamefont {E.~N.}\ \bibnamefont
  {Zalta}}}\ (\bibinfo {year} {2013})\ \bibinfo {edition} {fall 2013}\
  ed.\BibitemShut {Stop}%
\bibitem [{\citenamefont {Brukner}\ and\ \citenamefont
  {Zeilinger}(2003)}]{Brukner2003}%
  \BibitemOpen
  \bibfield  {author} {\bibinfo {author} {\bibfnamefont {C.}~\bibnamefont
  {Brukner}}\ and\ \bibinfo {author} {\bibfnamefont {A.}~\bibnamefont
  {Zeilinger}},\ }in\ \href@noop {} {\emph {\bibinfo {booktitle} {Time, Quantum
  and Information}}},\ \bibinfo {editor} {edited by\ \bibinfo {editor}
  {\bibfnamefont {L.}~\bibnamefont {Castell}}\ and\ \bibinfo {editor}
  {\bibfnamefont {O.}~\bibnamefont {Ischebeck}}}\ (\bibinfo  {publisher}
  {Springer},\ \bibinfo {year} {2003})\ \Eprint
  {http://arxiv.org/abs/quant-ph/0212084} {arXiv:quant-ph/0212084} \BibitemShut
  {NoStop}%
\bibitem [{\citenamefont {Fuchs}(2003)}]{Fuchs2003}%
  \BibitemOpen
  \bibfield  {author} {\bibinfo {author} {\bibfnamefont {C.~A.}\ \bibnamefont
  {Fuchs}},\ }\href {\doibase 10.1080/09500340308234548} {\bibfield  {journal}
  {\bibinfo  {journal} {J. Mod. Opt.}\ }\textbf {\bibinfo {volume} {50}},\
  \bibinfo {pages} {987} (\bibinfo {year} {2003})},\ \Eprint
  {http://arxiv.org/abs/quant-ph/0205039} {arXiv:quant-ph/0205039} \BibitemShut
  {NoStop}%
\bibitem [{\citenamefont {Spekkens}(2007)}]{toytheory}%
  \BibitemOpen
  \bibfield  {author} {\bibinfo {author} {\bibfnamefont {R.~W.}\ \bibnamefont
  {Spekkens}},\ }\href {\doibase 10.1103/PhysRevA.75.032110} {\bibfield
  {journal} {\bibinfo  {journal} {Phys. Rev. A}\ }\textbf {\bibinfo {volume}
  {75}},\ \bibinfo {pages} {032110} (\bibinfo {year} {2007})},\ \Eprint
  {http://arxiv.org/abs/quant-ph/0401052} {arXiv:quant-ph/0401052} \BibitemShut
  {NoStop}%
\bibitem [{\citenamefont {Fuchs}(2010{\natexlab{a}})}]{Fuchs2010}%
  \BibitemOpen
  \bibfield  {author} {\bibinfo {author} {\bibfnamefont {C.~A.}\ \bibnamefont
  {Fuchs}},\ }\href@noop {} {\enquote {\bibinfo {title} {{QB}ism, the perimeter
  of quantum bayesianism},}\ } (\bibinfo {year} {2010}{\natexlab{a}}),\ \Eprint
  {http://arxiv.org/abs/1003.5209} {arXiv:1003.5209} \BibitemShut {NoStop}%
\bibitem [{\citenamefont {Fuchs}(2010{\natexlab{b}})}]{Fuchs2010a}%
  \BibitemOpen
  \bibfield  {author} {\bibinfo {author} {\bibfnamefont {C.~A.}\ \bibnamefont
  {Fuchs}},\ }\href
  {http://www.cap.ca/en/article/quantum-bayesianism-perimeter} {\bibfield
  {journal} {\bibinfo  {journal} {Phys. Can.}\ }\textbf {\bibinfo {volume}
  {66}},\ \bibinfo {pages} {77} (\bibinfo {year} {2010}{\natexlab{b}})},\
  \Eprint {http://arxiv.org/abs/1003.5182} {arXiv:1003.5182} \BibitemShut
  {NoStop}%
\bibitem [{\citenamefont {Fuchs}\ \emph {et~al.}(2014)\citenamefont {Fuchs},
  \citenamefont {Mermin},\ and\ \citenamefont {Schack}}]{Fuchs2013}%
  \BibitemOpen
  \bibfield  {author} {\bibinfo {author} {\bibfnamefont {C.~A.}\ \bibnamefont
  {Fuchs}}, \bibinfo {author} {\bibfnamefont {N.~D.}\ \bibnamefont {Mermin}}, \
  and\ \bibinfo {author} {\bibfnamefont {R.}~\bibnamefont {Schack}},\ }\href
  {\doibase 10.1119/1.4874855} {\bibfield  {journal} {\bibinfo  {journal} {Am.
  J. Phys.}\ }\textbf {\bibinfo {volume} {82}},\ \bibinfo {pages} {749}
  (\bibinfo {year} {2014})},\ \Eprint {http://arxiv.org/abs/1311.5253}
  {arXiv:1311.5253} \BibitemShut {NoStop}%
\bibitem [{\citenamefont {Pusey}\ \emph {et~al.}(2012)\citenamefont {Pusey},
  \citenamefont {Barrett},\ and\ \citenamefont {Rudolph}}]{Pusey2012}%
  \BibitemOpen
  \bibfield  {author} {\bibinfo {author} {\bibfnamefont {M.~F.}\ \bibnamefont
  {Pusey}}, \bibinfo {author} {\bibfnamefont {J.}~\bibnamefont {Barrett}}, \
  and\ \bibinfo {author} {\bibfnamefont {T.}~\bibnamefont {Rudolph}},\ }\href
  {\doibase 10.1038/nphys2309} {\bibfield  {journal} {\bibinfo  {journal}
  {Nature Phys.}\ }\textbf {\bibinfo {volume} {8}},\ \bibinfo {pages} {475}
  (\bibinfo {year} {2012})},\ \Eprint {http://arxiv.org/abs/1111.3328}
  {arXiv:1111.3328} \BibitemShut {NoStop}%
\bibitem [{\citenamefont {Colbeck}\ and\ \citenamefont
  {Renner}(2012)}]{Colbeck2012}%
  \BibitemOpen
  \bibfield  {author} {\bibinfo {author} {\bibfnamefont {R.}~\bibnamefont
  {Colbeck}}\ and\ \bibinfo {author} {\bibfnamefont {R.}~\bibnamefont
  {Renner}},\ }\href {\doibase 10.1103/PhysRevLett.108.150402} {\bibfield
  {journal} {\bibinfo  {journal} {Phys. Rev. Lett.}\ }\textbf {\bibinfo
  {volume} {108}},\ \bibinfo {pages} {150402} (\bibinfo {year} {2012})},\
  \Eprint {http://arxiv.org/abs/1111.6597} {arXiv:1111.6597} \BibitemShut
  {NoStop}%
\bibitem [{\citenamefont {Colbeck}\ and\ \citenamefont
  {Renner}(2013)}]{Colbeck2013a}%
  \BibitemOpen
  \bibfield  {author} {\bibinfo {author} {\bibfnamefont {R.}~\bibnamefont
  {Colbeck}}\ and\ \bibinfo {author} {\bibfnamefont {R.}~\bibnamefont
  {Renner}},\ }\href@noop {} {\enquote {\bibinfo {title} {A system's wave
  function is uniquely determined by its underlying physical state},}\ }
  (\bibinfo {year} {2013}),\ \Eprint {http://arxiv.org/abs/1312.7353}
  {arXiv:1312.7353} \BibitemShut {NoStop}%
\bibitem [{\citenamefont {Aaronson}\ \emph {et~al.}(2013)\citenamefont
  {Aaronson}, \citenamefont {Bouland}, \citenamefont {Chua},\ and\
  \citenamefont {Lowther}}]{Aaronson2013}%
  \BibitemOpen
  \bibfield  {author} {\bibinfo {author} {\bibfnamefont {S.}~\bibnamefont
  {Aaronson}}, \bibinfo {author} {\bibfnamefont {A.}~\bibnamefont {Bouland}},
  \bibinfo {author} {\bibfnamefont {L.}~\bibnamefont {Chua}}, \ and\ \bibinfo
  {author} {\bibfnamefont {G.}~\bibnamefont {Lowther}},\ }\href {\doibase
  10.1103/PhysRevA.88.032111} {\bibfield  {journal} {\bibinfo  {journal} {Phys.
  Rev. A}\ }\textbf {\bibinfo {volume} {88}},\ \bibinfo {pages} {032111}
  (\bibinfo {year} {2013})},\ \Eprint {http://arxiv.org/abs/1303.2834}
  {arXiv:1303.2834} \BibitemShut {NoStop}%
\bibitem [{\citenamefont {Hardy}(2013)}]{Hardy2013}%
  \BibitemOpen
  \bibfield  {author} {\bibinfo {author} {\bibfnamefont {L.}~\bibnamefont
  {Hardy}},\ }\href {\doibase 10.1142/S0217979213450124} {\bibfield  {journal}
  {\bibinfo  {journal} {Int. J. Mod. Phys. B}\ }\textbf {\bibinfo {volume}
  {27}},\ \bibinfo {pages} {1345012} (\bibinfo {year} {2013})},\ \Eprint
  {http://arxiv.org/abs/1205.1439} {arXiv:1205.1439} \BibitemShut {NoStop}%
\bibitem [{\citenamefont {Patra}\ \emph {et~al.}(2013)\citenamefont {Patra},
  \citenamefont {Pironio},\ and\ \citenamefont {Massar}}]{Patra2013a}%
  \BibitemOpen
  \bibfield  {author} {\bibinfo {author} {\bibfnamefont {M.~K.}\ \bibnamefont
  {Patra}}, \bibinfo {author} {\bibfnamefont {S.}~\bibnamefont {Pironio}}, \
  and\ \bibinfo {author} {\bibfnamefont {S.}~\bibnamefont {Massar}},\ }\href
  {\doibase 10.1103/PhysRevLett.111.090402} {\bibfield  {journal} {\bibinfo
  {journal} {Phys. Rev. Lett.}\ }\textbf {\bibinfo {volume} {111}},\ \bibinfo
  {pages} {090402} (\bibinfo {year} {2013})},\ \Eprint
  {http://arxiv.org/abs/1211.1179} {arXiv:1211.1179} \BibitemShut {NoStop}%
\bibitem [{\citenamefont {Mansfield}(2014)}]{Mansfield2014}%
  \BibitemOpen
  \bibfield  {author} {\bibinfo {author} {\bibfnamefont {S.}~\bibnamefont
  {Mansfield}},\ }\href@noop {} {\enquote {\bibinfo {title} {Reality of the
  quantum state: A stronger psi-ontology theorem},}\ } (\bibinfo {year}
  {2014}),\ \Eprint {http://arxiv.org/abs/1412.0669} {arXiv:1412.0669}
  \BibitemShut {NoStop}%
\bibitem [{\citenamefont {Montina}(2015)}]{Montina2015}%
  \BibitemOpen
  \bibfield  {author} {\bibinfo {author} {\bibfnamefont {A.}~\bibnamefont
  {Montina}},\ }\href {\doibase 10.1142/S0217732315300013} {\bibfield
  {journal} {\bibinfo  {journal} {Mod. Phys. Lett. A}\ }\textbf {\bibinfo
  {volume} {30}},\ \bibinfo {pages} {1530001} (\bibinfo {year} {2015})},\
  \Eprint {http://arxiv.org/abs/1412.1723} {arXiv:1412.1723} \BibitemShut
  {NoStop}%
\bibitem [{\citenamefont {Leifer}(2014)}]{psireview}%
  \BibitemOpen
  \bibfield  {author} {\bibinfo {author} {\bibfnamefont {M.}~\bibnamefont
  {Leifer}},\ }\href {\doibase 10.12743/quanta.v3i1.22} {\bibfield  {journal}
  {\bibinfo  {journal} {Quanta}\ }\textbf {\bibinfo {volume} {3}},\ \bibinfo
  {pages} {67} (\bibinfo {year} {2014})},\ \Eprint
  {http://arxiv.org/abs/1409.1570} {arXiv:1409.1570} \BibitemShut {NoStop}%
\bibitem [{\citenamefont {Aharonov}\ and\ \citenamefont
  {Vaidman}(1993)}]{protect}%
  \BibitemOpen
  \bibfield  {author} {\bibinfo {author} {\bibfnamefont {Y.}~\bibnamefont
  {Aharonov}}\ and\ \bibinfo {author} {\bibfnamefont {L.}~\bibnamefont
  {Vaidman}},\ }\href {\doibase 10.1016/0375-9601(93)90724-E} {\bibfield
  {journal} {\bibinfo  {journal} {Phys. Lett. A}\ }\textbf {\bibinfo {volume}
  {178}},\ \bibinfo {pages} {38} (\bibinfo {year} {1993})},\ \Eprint
  {http://arxiv.org/abs/hep-th/9304147} {arXiv:hep-th/9304147} \BibitemShut
  {NoStop}%
\bibitem [{\citenamefont {Aharonov}\ \emph {et~al.}(1993)\citenamefont
  {Aharonov}, \citenamefont {Anandan},\ and\ \citenamefont
  {Vaidman}}]{meaning}%
  \BibitemOpen
  \bibfield  {author} {\bibinfo {author} {\bibfnamefont {Y.}~\bibnamefont
  {Aharonov}}, \bibinfo {author} {\bibfnamefont {J.}~\bibnamefont {Anandan}}, \
  and\ \bibinfo {author} {\bibfnamefont {L.}~\bibnamefont {Vaidman}},\ }\href
  {\doibase 10.1103/PhysRevA.47.4616} {\bibfield  {journal} {\bibinfo
  {journal} {Phys. Rev. A}\ }\textbf {\bibinfo {volume} {47}},\ \bibinfo
  {pages} {4616} (\bibinfo {year} {1993})}\BibitemShut {NoStop}%
\bibitem [{\citenamefont {Aharonov}\ \emph {et~al.}(1996)\citenamefont
  {Aharonov}, \citenamefont {Anandan},\ and\ \citenamefont {Vaidman}}]{again}%
  \BibitemOpen
  \bibfield  {author} {\bibinfo {author} {\bibfnamefont {Y.}~\bibnamefont
  {Aharonov}}, \bibinfo {author} {\bibfnamefont {J.}~\bibnamefont {Anandan}}, \
  and\ \bibinfo {author} {\bibfnamefont {L.}~\bibnamefont {Vaidman}},\ }\href
  {\doibase 10.1007/BF02058891} {\bibfield  {journal} {\bibinfo  {journal}
  {Found. Phys.}\ }\textbf {\bibinfo {volume} {26}},\ \bibinfo {pages} {117}
  (\bibinfo {year} {1996})},\ \Eprint {http://arxiv.org/abs/hep-th/9408153}
  {arXiv:hep-th/9408153} \BibitemShut {NoStop}%
\bibitem [{\citenamefont {Gao}(2013)}]{shan}%
  \BibitemOpen
  \bibfield  {author} {\bibinfo {author} {\bibfnamefont {S.}~\bibnamefont
  {Gao}},\ }\href@noop {} {\enquote {\bibinfo {title} {Distinct quantum states
  cannot be compatible with a single state of reality},}\ } (\bibinfo {year}
  {2013}),\ \bibinfo {note}
  {\href{http://philsci-archive.pitt.edu/9609/}{PhilSci:9609}}\BibitemShut
  {NoStop}%
\bibitem [{\citenamefont {Vaidman}(2014)}]{lev}%
  \BibitemOpen
  \bibfield  {author} {\bibinfo {author} {\bibfnamefont {L.}~\bibnamefont
  {Vaidman}},\ }in\ \href@noop {} {\emph {\bibinfo {booktitle} {Protective
  Measurement and Quantum Reality: Towards a New Understanding of Quantum
  Mechanics}}},\ \bibinfo {editor} {edited by\ \bibinfo {editor} {\bibfnamefont
  {S.}~\bibnamefont {Gao}}}\ (\bibinfo  {publisher} {Cambridge University
  Press},\ \bibinfo {year} {2014})\ pp.\ \bibinfo {pages} {15--27},\ \Eprint
  {http://arxiv.org/abs/1401.6696} {arXiv:1401.6696} \BibitemShut {NoStop}%
\bibitem [{\citenamefont {Hetzroni}\ and\ \citenamefont
  {Rohrlich}(2014)}]{hetroh}%
  \BibitemOpen
  \bibfield  {author} {\bibinfo {author} {\bibfnamefont {G.}~\bibnamefont
  {Hetzroni}}\ and\ \bibinfo {author} {\bibfnamefont {D.}~\bibnamefont
  {Rohrlich}},\ }in\ \href@noop {} {\emph {\bibinfo {booktitle} {Protective
  Measurement and Quantum Reality: Towards a New Understanding of Quantum
  Mechanics}}},\ \bibinfo {editor} {edited by\ \bibinfo {editor} {\bibfnamefont
  {S.}~\bibnamefont {Gao}}}\ (\bibinfo  {publisher} {Cambridge University
  Press},\ \bibinfo {year} {2014})\ pp.\ \bibinfo {pages} {135--144},\ \Eprint
  {http://arxiv.org/abs/1403.1590} {arXiv:1403.1590} \BibitemShut {NoStop}%
\bibitem [{\citenamefont {Gao}(2015)}]{Gao2015}%
  \BibitemOpen
  \bibfield  {author} {\bibinfo {author} {\bibfnamefont {S.}~\bibnamefont
  {Gao}},\ }\href {\doibase 10.1016/j.shpsb.2015.07.006} {\bibfield  {journal}
  {\bibinfo  {journal} {Stud. Hist. Phil. Mod. Phys.}\ } (\bibinfo {year}
  {2015}),\ 10.1016/j.shpsb.2015.07.006},\ \bibinfo {note} {in press.},\
  \Eprint {http://arxiv.org/abs/arXiv:1508.07684} {arXiv:1508.07684}
  \BibitemShut {NoStop}%
\bibitem [{\citenamefont {Unruh}(1994)}]{unruh}%
  \BibitemOpen
  \bibfield  {author} {\bibinfo {author} {\bibfnamefont {W.}~\bibnamefont
  {Unruh}},\ }\href {\doibase 10.1103/PhysRevA.50.882} {\bibfield  {journal}
  {\bibinfo  {journal} {Phys. Rev. A}\ }\textbf {\bibinfo {volume} {50}},\
  \bibinfo {pages} {882} (\bibinfo {year} {1994})},\ \Eprint
  {http://arxiv.org/abs/hep-th/9308061} {arXiv:hep-th/9308061} \BibitemShut
  {NoStop}%
\bibitem [{\citenamefont {Rovelli}(1994)}]{rovelli}%
  \BibitemOpen
  \bibfield  {author} {\bibinfo {author} {\bibfnamefont {C.}~\bibnamefont
  {Rovelli}},\ }\href {\doibase 10.1103/PhysRevA.50.2788} {\bibfield  {journal}
  {\bibinfo  {journal} {Phys. Rev. A}\ }\textbf {\bibinfo {volume} {50}},\
  \bibinfo {pages} {2788} (\bibinfo {year} {1994})}\BibitemShut {NoStop}%
\bibitem [{\citenamefont {Ghose}\ and\ \citenamefont {Home}(1995)}]{Ghose1995}%
  \BibitemOpen
  \bibfield  {author} {\bibinfo {author} {\bibfnamefont {P.}~\bibnamefont
  {Ghose}}\ and\ \bibinfo {author} {\bibfnamefont {D.}~\bibnamefont {Home}},\
  }\href {\doibase 10.1007/BF02059528} {\bibfield  {journal} {\bibinfo
  {journal} {Found. Phys.}\ }\textbf {\bibinfo {volume} {25}},\ \bibinfo
  {pages} {1105} (\bibinfo {year} {1995})}\BibitemShut {NoStop}%
\bibitem [{\citenamefont {Uffink}(1999)}]{uffink}%
  \BibitemOpen
  \bibfield  {author} {\bibinfo {author} {\bibfnamefont {J.}~\bibnamefont
  {Uffink}},\ }\href {\doibase 10.1103/PhysRevA.60.3474} {\bibfield  {journal}
  {\bibinfo  {journal} {Phys. Rev. A}\ }\textbf {\bibinfo {volume} {60}},\
  \bibinfo {pages} {3474} (\bibinfo {year} {1999})},\ \Eprint
  {http://arxiv.org/abs/quant-ph/9903007} {arXiv:quant-ph/9903007} \BibitemShut
  {NoStop}%
\bibitem [{\citenamefont {D'Ariano}\ and\ \citenamefont
  {Yuen}(1996)}]{dariano}%
  \BibitemOpen
  \bibfield  {author} {\bibinfo {author} {\bibfnamefont {G.}~\bibnamefont
  {D'Ariano}}\ and\ \bibinfo {author} {\bibfnamefont {H.}~\bibnamefont
  {Yuen}},\ }\href {\doibase 10.1103/PhysRevLett.76.2832} {\bibfield  {journal}
  {\bibinfo  {journal} {Phys. Rev. Lett.}\ }\textbf {\bibinfo {volume} {76}},\
  \bibinfo {pages} {2832} (\bibinfo {year} {1996})}\BibitemShut {NoStop}%
\bibitem [{\citenamefont {Hari~Dass}\ and\ \citenamefont
  {Qureshi}(1999)}]{dass}%
  \BibitemOpen
  \bibfield  {author} {\bibinfo {author} {\bibfnamefont {N.}~\bibnamefont
  {Hari~Dass}}\ and\ \bibinfo {author} {\bibfnamefont {T.}~\bibnamefont
  {Qureshi}},\ }\href {\doibase 10.1103/PhysRevA.59.2590} {\bibfield  {journal}
  {\bibinfo  {journal} {Phys. Rev. A}\ }\textbf {\bibinfo {volume} {59}},\
  \bibinfo {pages} {2590} (\bibinfo {year} {1999})},\ \Eprint
  {http://arxiv.org/abs/quant-ph/9805012} {arXiv:quant-ph/9805012} \BibitemShut
  {NoStop}%
\bibitem [{\citenamefont {Uffink}(2013)}]{uffink_v_shan}%
  \BibitemOpen
  \bibfield  {author} {\bibinfo {author} {\bibfnamefont {J.}~\bibnamefont
  {Uffink}},\ }\href {\doibase 10.1016/j.shpsb.2013.07.002} {\bibfield
  {journal} {\bibinfo  {journal} {Stud. Hist. Phil. Mod. Phys.}\ }\textbf
  {\bibinfo {volume} {44}},\ \bibinfo {pages} {519 } (\bibinfo {year}
  {2013})},\ \bibinfo {note}
  {\href{http://philsci-archive.pitt.edu/9286/}{PhilSci:9286}}\BibitemShut
  {NoStop}%
\bibitem [{\citenamefont {Hagar}(2014)}]{hagar}%
  \BibitemOpen
  \bibfield  {author} {\bibinfo {author} {\bibfnamefont {A.}~\bibnamefont
  {Hagar}},\ }\href {http://philpapers.org/rec/HAGWPM} {\enquote {\bibinfo
  {title} {Does protective measurement tell us anything about quantum
  reality?}}\ } (\bibinfo {year} {2014})\BibitemShut {NoStop}%
\bibitem [{\citenamefont {Schlosshauer}\ and\ \citenamefont
  {Claringbold}(2014)}]{schcla}%
  \BibitemOpen
  \bibfield  {author} {\bibinfo {author} {\bibfnamefont {M.}~\bibnamefont
  {Schlosshauer}}\ and\ \bibinfo {author} {\bibfnamefont {T.~V.~B.}\
  \bibnamefont {Claringbold}},\ }in\ \href@noop {} {\emph {\bibinfo {booktitle}
  {Protective Measurement and Quantum Reality: Towards a New Understanding of
  Quantum Mechanics}}},\ \bibinfo {editor} {edited by\ \bibinfo {editor}
  {\bibfnamefont {S.}~\bibnamefont {Gao}}}\ (\bibinfo  {publisher} {Cambridge
  University Press},\ \bibinfo {year} {2014})\ pp.\ \bibinfo {pages}
  {180--194},\ \Eprint {http://arxiv.org/abs/1402.1217} {arXiv:1402.1217}
  \BibitemShut {NoStop}%
\bibitem [{\citenamefont {Einstein}\ \emph {et~al.}(1935)\citenamefont
  {Einstein}, \citenamefont {Podolsky},\ and\ \citenamefont {Rosen}}]{epr}%
  \BibitemOpen
  \bibfield  {author} {\bibinfo {author} {\bibfnamefont {A.}~\bibnamefont
  {Einstein}}, \bibinfo {author} {\bibfnamefont {B.}~\bibnamefont {Podolsky}},
  \ and\ \bibinfo {author} {\bibfnamefont {N.}~\bibnamefont {Rosen}},\ }\href
  {\doibase 10.1103/PhysRev.47.777} {\bibfield  {journal} {\bibinfo  {journal}
  {Phys. Rev.}\ }\textbf {\bibinfo {volume} {47}},\ \bibinfo {pages} {777}
  (\bibinfo {year} {1935})}\BibitemShut {NoStop}%
\bibitem [{\citenamefont {Misra}\ and\ \citenamefont {Sudarshan}(1977)}]{zeno}%
  \BibitemOpen
  \bibfield  {author} {\bibinfo {author} {\bibfnamefont {B.}~\bibnamefont
  {Misra}}\ and\ \bibinfo {author} {\bibfnamefont {E.~C.~G.}\ \bibnamefont
  {Sudarshan}},\ }\href {\doibase 10.1063/1.523304} {\bibfield  {journal}
  {\bibinfo  {journal} {J. Math. Phys.}\ }\textbf {\bibinfo {volume} {18}},\
  \bibinfo {pages} {756} (\bibinfo {year} {1977})}\BibitemShut {NoStop}%
\bibitem [{\citenamefont {Chuang}\ and\ \citenamefont
  {Nielsen}(1997)}]{processtomo}%
  \BibitemOpen
  \bibfield  {author} {\bibinfo {author} {\bibfnamefont {I.~L.}\ \bibnamefont
  {Chuang}}\ and\ \bibinfo {author} {\bibfnamefont {M.~A.}\ \bibnamefont
  {Nielsen}},\ }\href {\doibase 10.1080/09500349708231894} {\bibfield
  {journal} {\bibinfo  {journal} {J. Mod. Opt.}\ }\textbf {\bibinfo {volume}
  {44}},\ \bibinfo {pages} {2455} (\bibinfo {year} {1997})},\ \Eprint
  {http://arxiv.org/abs/quant-ph/9610001} {arXiv:quant-ph/9610001} \BibitemShut
  {NoStop}%
\bibitem [{\citenamefont {Kribs}(2003)}]{kribs}%
  \BibitemOpen
  \bibfield  {author} {\bibinfo {author} {\bibfnamefont {D.~W.}\ \bibnamefont
  {Kribs}},\ }\href {\doibase 10.1017/S0013091501000980} {\bibfield  {journal}
  {\bibinfo  {journal} {Proc. Edinb. Math. Soc.}\ }\textbf {\bibinfo {volume}
  {46}},\ \bibinfo {pages} {421} (\bibinfo {year} {2003})},\ \Eprint
  {http://arxiv.org/abs/math/0309390} {arXiv:math/0309390} \BibitemShut
  {NoStop}%
\bibitem [{\citenamefont {Wiesner}(1983)}]{money}%
  \BibitemOpen
  \bibfield  {author} {\bibinfo {author} {\bibfnamefont {S.}~\bibnamefont
  {Wiesner}},\ }\href {\doibase 10.1145/1008908.1008920} {\bibfield  {journal}
  {\bibinfo  {journal} {SIGACT News}\ }\textbf {\bibinfo {volume} {15}},\
  \bibinfo {pages} {78} (\bibinfo {year} {1983})}\BibitemShut {NoStop}%
\bibitem [{\citenamefont {{Brodutch}}\ \emph {et~al.}(2014)\citenamefont
  {{Brodutch}}, \citenamefont {{Nagaj}}, \citenamefont {{Sattath}},\ and\
  \citenamefont {{Unruh}}}]{forgery}%
  \BibitemOpen
  \bibfield  {author} {\bibinfo {author} {\bibfnamefont {A.}~\bibnamefont
  {{Brodutch}}}, \bibinfo {author} {\bibfnamefont {D.}~\bibnamefont {{Nagaj}}},
  \bibinfo {author} {\bibfnamefont {O.}~\bibnamefont {{Sattath}}}, \ and\
  \bibinfo {author} {\bibfnamefont {D.}~\bibnamefont {{Unruh}}},\ }\href@noop
  {} {\enquote {\bibinfo {title} {{An adaptive attack on Wiesner's quantum
  money}},}\ } (\bibinfo {year} {2014}),\ \Eprint
  {http://arxiv.org/abs/1404.1507} {arXiv:1404.1507} \BibitemShut {NoStop}%
\bibitem [{\citenamefont {Gutoski}\ and\ \citenamefont
  {Johnston}(2014)}]{unitarytomo}%
  \BibitemOpen
  \bibfield  {author} {\bibinfo {author} {\bibfnamefont {G.}~\bibnamefont
  {Gutoski}}\ and\ \bibinfo {author} {\bibfnamefont {N.}~\bibnamefont
  {Johnston}},\ }\href {\doibase 10.1063/1.4867625} {\bibfield  {journal}
  {\bibinfo  {journal} {J. Math. Phys.}\ }\textbf {\bibinfo {volume} {55}},\
  \bibinfo {pages} {032201} (\bibinfo {year} {2014})},\ \Eprint
  {http://arxiv.org/abs/1309.0840} {arXiv:1309.0840} \BibitemShut {NoStop}%
\bibitem [{\citenamefont {James}\ and\ \citenamefont {Kosut}(2010)}]{qec}%
  \BibitemOpen
  \bibfield  {author} {\bibinfo {author} {\bibfnamefont {M.~R.}\ \bibnamefont
  {James}}\ and\ \bibinfo {author} {\bibfnamefont {R.~L.}\ \bibnamefont
  {Kosut}},\ }in\ \href@noop {} {\emph {\bibinfo {booktitle} {The Control
  Handbook: Control System Applications}}},\ \bibinfo {editor} {edited by\
  \bibinfo {editor} {\bibfnamefont {W.~S.}\ \bibnamefont {Levine}}}\ (\bibinfo
  {publisher} {Taylor and Francis},\ \bibinfo {year} {2010})\ Chap.~\bibinfo
  {chapter} {31}\BibitemShut {NoStop}%
\bibitem [{\citenamefont {Bartlett}\ \emph {et~al.}(2012)\citenamefont
  {Bartlett}, \citenamefont {Rudolph},\ and\ \citenamefont {Spekkens}}]{erl}%
  \BibitemOpen
  \bibfield  {author} {\bibinfo {author} {\bibfnamefont {S.~D.}\ \bibnamefont
  {Bartlett}}, \bibinfo {author} {\bibfnamefont {T.}~\bibnamefont {Rudolph}}, \
  and\ \bibinfo {author} {\bibfnamefont {R.~W.}\ \bibnamefont {Spekkens}},\
  }\href {\doibase 10.1103/PhysRevA.86.012103} {\bibfield  {journal} {\bibinfo
  {journal} {Phys. Rev. A}\ }\textbf {\bibinfo {volume} {86}},\ \bibinfo
  {pages} {012103} (\bibinfo {year} {2012})},\ \Eprint
  {http://arxiv.org/abs/1111.5057} {arXiv:1111.5057} \BibitemShut {NoStop}%
\bibitem [{\citenamefont {Karanjai}\ \emph {et~al.}(2015)\citenamefont
  {Karanjai}, \citenamefont {Cavalcanti}, \citenamefont {Bartlett},\ and\
  \citenamefont {Rudolph}}]{Karanjai}%
  \BibitemOpen
  \bibfield  {author} {\bibinfo {author} {\bibfnamefont {A.}~\bibnamefont
  {Karanjai}}, \bibinfo {author} {\bibfnamefont {E.~G.}\ \bibnamefont
  {Cavalcanti}}, \bibinfo {author} {\bibfnamefont {S.~D.}\ \bibnamefont
  {Bartlett}}, \ and\ \bibinfo {author} {\bibfnamefont {T.}~\bibnamefont
  {Rudolph}},\ }\href {http://dx.doi.org/10.1088/1367-2630/17/7/073015}
  {\bibfield  {journal} {\bibinfo  {journal} {New J. Phys.}\ }\textbf {\bibinfo
  {volume} {17}},\ \bibinfo {pages} {073015} (\bibinfo {year} {2015})},\
  \Eprint {http://arxiv.org/abs/1503.05203} {arXiv:1503.05203} \BibitemShut
  {NoStop}%
\bibitem [{\citenamefont {Caves}\ \emph {et~al.}(2002)\citenamefont {Caves},
  \citenamefont {Fuchs},\ and\ \citenamefont {Schack}}]{statedefinetti}%
  \BibitemOpen
  \bibfield  {author} {\bibinfo {author} {\bibfnamefont {C.~M.}\ \bibnamefont
  {Caves}}, \bibinfo {author} {\bibfnamefont {C.~A.}\ \bibnamefont {Fuchs}}, \
  and\ \bibinfo {author} {\bibfnamefont {R.}~\bibnamefont {Schack}},\ }\href
  {\doibase http://dx.doi.org/10.1063/1.1494475} {\bibfield  {journal}
  {\bibinfo  {journal} {J. Math. Phys.}\ }\textbf {\bibinfo {volume} {43}},\
  \bibinfo {pages} {4537} (\bibinfo {year} {2002})},\ \Eprint
  {http://arxiv.org/abs/quant-ph/0104088} {arXiv:quant-ph/0104088} \BibitemShut
  {NoStop}%
\bibitem [{\citenamefont {Ferrie}\ and\ \citenamefont
  {Combes}(2014)}]{FerrieCombes14}%
  \BibitemOpen
  \bibfield  {author} {\bibinfo {author} {\bibfnamefont {C.}~\bibnamefont
  {Ferrie}}\ and\ \bibinfo {author} {\bibfnamefont {J.}~\bibnamefont
  {Combes}},\ }\href {\doibase 10.1103/PhysRevLett.113.120404} {\bibfield
  {journal} {\bibinfo  {journal} {Phys. Rev. Lett.}\ }\textbf {\bibinfo
  {volume} {113}},\ \bibinfo {pages} {120404} (\bibinfo {year} {2014})},\
  \Eprint {http://arxiv.org/abs/1403.2362} {arXiv:1403.2362} \BibitemShut
  {NoStop}%
\bibitem [{\citenamefont {Ipsen}(2015)}]{Ipsen15}%
  \BibitemOpen
  \bibfield  {author} {\bibinfo {author} {\bibfnamefont {A.~C.}\ \bibnamefont
  {Ipsen}},\ }\href {\doibase 10.1103/PhysRevA.91.062120} {\bibfield  {journal}
  {\bibinfo  {journal} {Phys. Rev. A}\ }\textbf {\bibinfo {volume} {91}},\
  \bibinfo {pages} {062120} (\bibinfo {year} {2015})},\ \Eprint
  {http://arxiv.org/abs/1409.3538} {arXiv:1409.3538} \BibitemShut {NoStop}%
\bibitem [{\citenamefont {Pusey}(2014)}]{Pusey14}%
  \BibitemOpen
  \bibfield  {author} {\bibinfo {author} {\bibfnamefont {M.~F.}\ \bibnamefont
  {Pusey}},\ }\href {\doibase 10.1103/PhysRevLett.113.200401} {\bibfield
  {journal} {\bibinfo  {journal} {Phys. Rev. Lett.}\ }\textbf {\bibinfo
  {volume} {113}},\ \bibinfo {pages} {200401} (\bibinfo {year} {2014})},\
  \Eprint {http://arxiv.org/abs/1409.1535} {arXiv:1409.1535} \BibitemShut
  {NoStop}%
\end{thebibliography}%

\end{document}